\documentclass[aps,prx,reprint,preprintnumbers,superscriptaddress,nofootinbib,longbibliography,floatfix]{revtex4-1}

\pdfoutput=1

\usepackage{graphicx, subcaption}
\usepackage{dcolumn}
\usepackage{bm}
\usepackage{xcolor}
\usepackage{amsmath}
\usepackage{multirow}
\usepackage{caption}
\usepackage{float}
\usepackage{hyperref}
\hypersetup{
  colorlinks=true,
  citecolor=blue,
  linkcolor=blue,
  urlcolor=blue
}

\DeclareRobustCommand{\Sec}[1]{Sec.~\ref{#1}}

\DeclareRobustCommand{\App}[1]{App.~\ref{#1}}
\DeclareRobustCommand{\Tab}[1]{Table~\ref{#1}}

\DeclareRobustCommand{\Fig}[1]{Fig.~\ref{#1}}

\DeclareRobustCommand{\Ref}[1]{Ref.~\cite{#1}}

\newcommand\DoubleImageWidth{0.6}
\newcommand\SingleImageWidth{1}

\begin{document}


\title{Self-supervised Anomaly Detection for New Physics\
}

\author{Barry M. Dillon}
 \email{dillon@thphys.uni-heidelberg.de}
\affiliation{%
Universit\"at Heidelberg, Heidelberg, Germany
}%

\author{Radha Mastandrea}
\email{rmastand@berkeley.edu}
\affiliation{Department of Physics, University of California, Berkeley, CA 94720, USA}
\affiliation{Physics Division, Lawrence Berkeley National Laboratory, Berkeley, CA 94720, USA}

\author{Benjamin Nachman}
\email{bpnachman@lbl.gov}
\affiliation{Physics Division, Lawrence Berkeley National Laboratory, Berkeley, CA 94720, USA}
\affiliation{Berkeley Institute for Data Science, University of California, Berkeley, CA 94720, USA}

\date{\today}

\begin{abstract}
We investigate a method of model-agnostic anomaly detection through studying jets, collimated sprays of particles produced in high-energy collisions. We train a transformer neural network to encode simulated QCD ``event space" dijets into a low-dimensional ``latent space" representation. We optimize the network using the self-supervised contrastive loss, which encourages the preservation of known physical symmetries of the dijets. We then train a binary classifier to discriminate a BSM resonant dijet signal from a QCD dijet background both in the event space and the latent space representations. We find the classifier performances on the event and latent spaces to be comparable. We finally perform an anomaly detection search using a weakly supervised bump hunt on the latent space dijets, finding again a comparable performance to a search run on the physical space dijets. This opens the door to using low-dimensional latent representations as a computationally efficient space for resonant anomaly detection in generic particle collision events.

\end{abstract}

\maketitle


\section{\label{sec:level1}Introduction}

A central goal of high energy physics is to find the theory that will supersede the Standard Model. A number of competing models exist, with many involving new resonant particles \cite{Craig:2016rqv, Kim:2019rhy} such as supersymmetric partners or weakly interacting massive particles. However, the number of candidate particles is too large to justify a hunt-and-pick procedure of data analysis. Therefore it is advantageous to consider methods of anomaly detection that are agnostic to a particular underlying model of new physics.

There are a number of recent machine learning (ML) based anomaly detection proposals designed to reduce model dependence (see Refs.~\cite{Kasieczka:2021xcg,Aarrestad:2021oeb,Karagiorgi:2021ngt,Feickert:2021ajf} for overviews). Notably, most existing methods (including the first result with data from ATLAS~\cite{collaboration2020dijet}) in the field are best-performing in low-dimensional spaces. However, a single event from a particle collision experiment can have on the order of a thousand degrees of freedom. 

A resolution to this tension is to reduce the dimensionality of an entire particle collision event while preserving its essential character such that a search for anomalies can be done in the reduced-dimension space.  A number of methods exist for carrying out this phase space reduction. For example, one could choose a set of observables (e.g. mass, multiplicity) and perform anomaly detection on this set. However, attempts to reduce dimensionality through selecting a choice of observables implicitly favors a certain class of models. 

Dimensionality reduction can also be performed with unsupervised ML techniques, which ensure a model-agnostic approach.  A common tool in the ML literature for this compression is the autoencoder (AE).  An autoencoder is a pair of neural networks whereby one function encodes \textit{event space} data into a \textit{latent space} and a second function decodes the latent space back into the event space.  No labels are needed because the AE is trained to ensure that the composition of the encoder and decoder is close to the identity to produce high reconstruction efficiency.  While effective (see e.g.~\cite{Hajer:2018kqm,Farina:2018fyg,Heimel:2018mkt, Bortolato:2021zic}), AE-based tools may not be ideal for anomaly detection. For one, there is nothing in their architectures or loss functions that ensure that anomalies, or basic physical (i.e. geometric) properties, of events are preserved by the encoder. Additionally, most AE studied so far cannot process a variable number of inputs per event, which would require the decoder to generate a variable number of outputs and the loss to compare events with a variable dimensionality.

These problems can be circumvented by making use of self-supervised contrastive learning techniques, which use only the inherent symmetries of physical data to perform a dimensionality reduction. Recent studies have explored this; for example with astronomical images in \Ref{https://doi.org/10.48550/arxiv.2012.13083}, and with constituent level jet data in \Ref{Dillon:2021gag}.  In the former, a ResNet50 architecture~\cite{https://doi.org/10.48550/arxiv.1512.03385} is used to map each astronomical image to a latent representation, while in the latter, a permutation-invariant transformer-encoder architecture maps jet constituents to a latent representation. 
The networks are trained on the contrastive loss, which ensures that the latent space representations faithfully model the physical symmetries of the original objects. 
Further analysis is then done directly on the latent space representations.

\begin{figure*}
    \centering
    \includegraphics[width=\SingleImageWidth\linewidth]{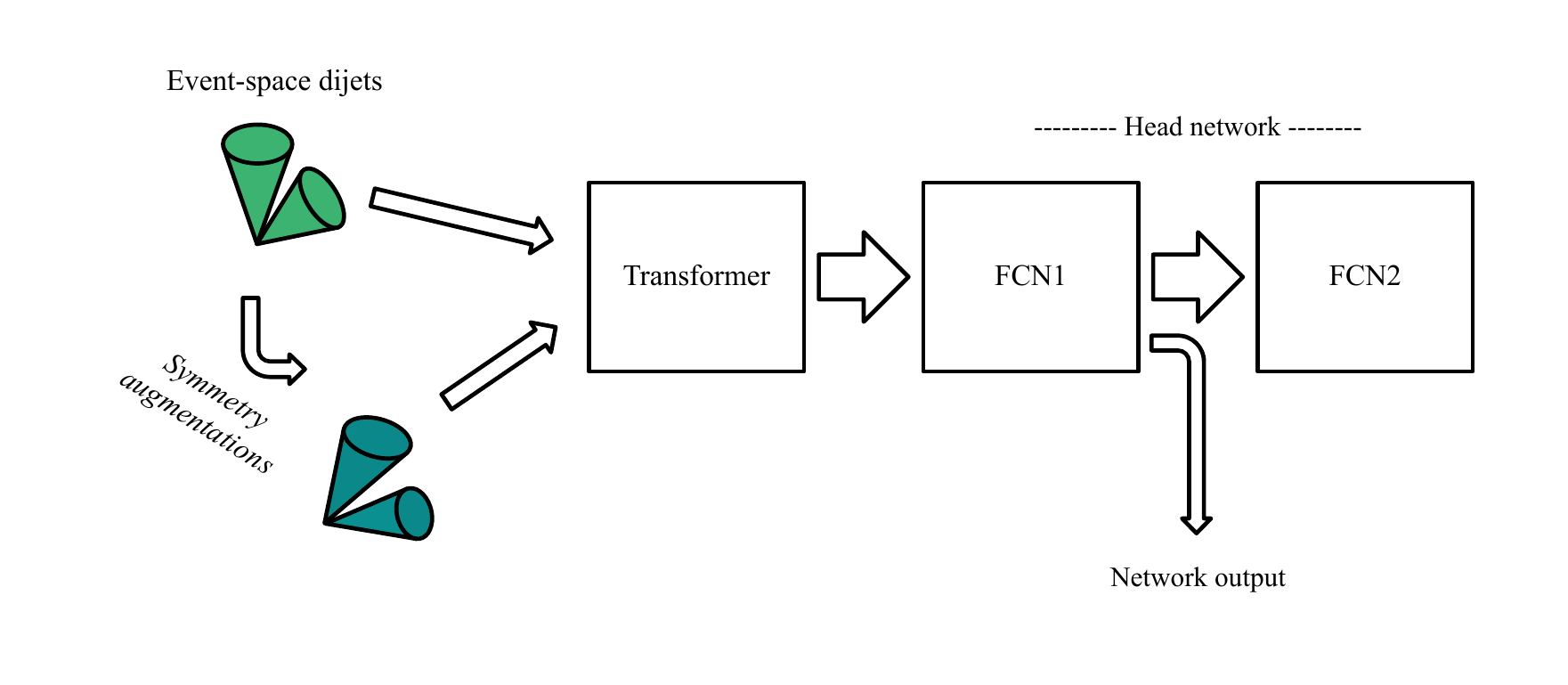}
    \caption{A schematic of the full transformer-encoder network. Event space dijets and their symmetry-augmented versions are fed as input into the network, which creates a mapping into the latent space by training on the contrastive loss function. The output of the transformer-encoder network is then passed through a head network, consisting of two fully connected layers (FCN1 and FCN2). In practice, the representations from the first fully connected layer perform the best in signal versus background classification tasks.}
    \label{fig:network_schematic}
\end{figure*}

In this paper, we continue the explorations of latent space representations of particle collisions originally carried out in \Ref{Dillon:2021gag}. We first demonstrate that particle collisions can be well-modeled in a latent space representation with a dimensionality that is an order of magnitude smaller than that of the original events. As part of this work, we extend the per-jet work of \Ref{Dillon:2021gag} to a per-event structure.  We then conduct a low-dimension model-agnostic anomaly search in the latent space representations of the particle collision events.
For this we use the Classification Without Labels (CWoLa) technique \cite{Metodiev:2017vrx,Collins:2018epr,Collins:2019jip}, which uses deep neural-network classifiers to distinguish between anomaly-enriched events and anomaly-depleted events.  We conduct these studies on dijet resonance events in the LHC Olympics dataset \cite{LHCOlympics}.

The structure of this paper is as follows. In \Sec{sec:methods}, we motivate the relevance of contrastive learning to modeling particle collisions and introduce a dataset of dijet events. We further outline a set of \textit{symmetry augmentations} for the contrastive loss function that leave the essential character of dijet events invariant and explain how these are used in the contrastive learning approach.
Lastly, in this section we outline the CWoLa anomaly detection method where we will use the self-supervised event representations.
In \Sec{sec:latent_space}, we implement the contrastive learning method using a transformer neural network to map the dijet events from the event space into a latent space and evaluate the efficiency of the encoding.
In \Sec{sec:cwola}, we use the CWoLa method to perform a relevant, but simplified, anomaly bump-hunt analysis using the latent space representations for the dijet events.  The paper ends with conclusions and outlook in Sec.~\ref{sec:conclusions}.

\vspace{20mm}

\section{\label{sec:methods}Methods}

Our overarching goal is to optimize a mapping from the event space of particle collision events (i.e. the representation in the space of the individual particles) to a new latent space representation. The mapping between the event and latent space representations of particle collisions should maximally exploit the physical symmetries of particle collision events. In this way, the dimensionality of the events can be reduced from a few hundred degrees of freedom (corresponding to the momentum 4-vectors of the particles) to a few tens.

Such a mapping can be realised by using a \textit{transformer-encoder} neural network architecture \cite{chen2020simple}. Event space events are fed into a transformer neural network where they are embedded into a reduced-dimension latent space. A distinguishing feature of the transformer architecture is permutation invariance: the latent space representations are invariant with respect to the order that the event constituents are fed into the network. The transformer-encoder network consists of 4 heads each with 2 layers.  The output of the transformer-encoder network is fed into a two layers of fully connected networks of the same latent space size. These architecture parameters were not heavily optimized.

See \Fig{fig:network_schematic} for a schematic of the full network. We find, as do the authors of \Ref{chen2020simple} and \Ref{Dillon:2021gag}, that the latent space representation of the first head layer output gives a better representation than that of the final output layer, or that of the transformer-encoder output.

\subsection{\label{sec:data_prep}Data Selection and Preparation}

For this paper, we focus on the LHC 2020 Olympics R\&D dataset \cite{LHCOlympics,Kasieczka:2021xcg}. The full dataset consists of 1,000,000 background dijet events (Standard
Model Quantum Chromodynamic (QCD) dijets) and 100,000 signal dijet events.
The signal comes from the process $Z'\rightarrow X(\rightarrow
q\overline{q})Y(\rightarrow q\overline{q})$, with three new resonances $Z'$ ($3.5$~TeV), $X$ (500~GeV),
and $Y$ (100~GeV).
The only trigger is a single large-radius jet ($R=1$) trigger with a $p_T$ threshold of $1.2$ TeV.  The events are generated with \texttt{Pythia}~8.219~\cite{Sjostrand:2006za,Sjostrand:2014zea} and \texttt{Delphes}~3.4.1~\cite{deFavereau:2013fsa}. Each event contains up to 700 particles with three degrees of freedom (DoF) $p_T$, $\eta$, $\phi$. The average number of nonzero DoF per event is 506 $\pm$ 174. 

For each event, we cluster the jets using \texttt{FastJet} \cite{Cacciari:2011ma,Cacciari:2005hq} with a radius $R = 0.8$. We select the two
highest-mass jets from each event, and we select the 50 hardest (highest
$p_T$) constituents from each event, zero-padding for any jet with fewer than 50. Note that the average number of constituents per event is 81 with a standard deviation of 16.15. However, we found that including more than 50 constituents per jet did not lead to an appreciable improvement in the performance of the transformer-encoder network.
The constituents are assumed to be massless, and so the relevant degrees of freedom for each constituent are $(p_T,\eta,\phi)$. 

For analysis, we select jets from the windows $p_T \in [800, 3000]$ GeV and $\eta \in [-3, 3]$. The cut on $p_T$ was chosen such that invariant dijet mass
$m_{JJ}$ has a lower bound at approximately 2 TeV. This cut removes approximately 12\% of eligible events from the LHCO dataset and has the benefit of removing a small tail of events with $m_{JJ}$ below 2 TeV that could appear to be artificially anomalous to the transformer-encoder network.

\begin{figure}
    \begin{center}
    \includegraphics[width=\DoubleImageWidth\linewidth]{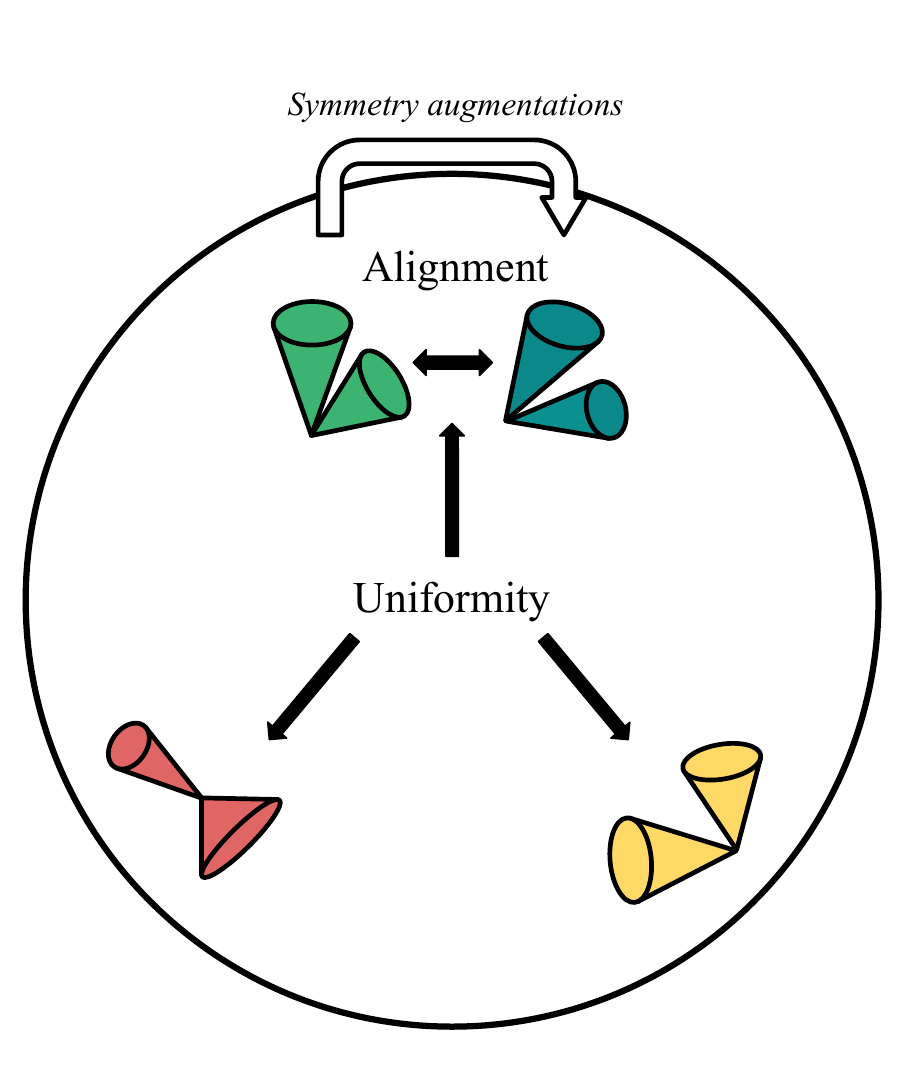}
    \end{center}
    \caption{An illustration of the latent space of the dijet events, built by a transformer-encoder network trained on the contrastive loss. The loss function optimizes for both alignment of dijets and their symmetry-augmented versions, and uniformity of physically distinct dijets.}
    \label{fig:latent_schematic}
\end{figure}

\subsection{\label{sec:clr}Contrastive  Learning }

The contrastive learning method is self-supervised, meaning that it is trained using ``pseudo-labels" rather than truth labels.
Supervised approaches use truth labels which exactly identify the truth label of the data.
Pseudo-labels are artificial labels created from the data alone, without access to the truth labels.
This means that the contrastive learning method is also unsupervised and receives no information as to whether the training samples are signal or background. 
Following JetCLR \cite{Dillon:2021gag}, the pseudo-labels are used to identify jets which are related to each other via some augmentation, for example a symmetry transformation.
Using the pseudo-labels, this technique aims to construct a latent space representation of events that exploits their physical symmetries.

As an example, consider the transformer-encoder network's encoding of a dijet event $\boldsymbol{r_{j}}$, and the encoding of an \textit{augmented} version of that event $\boldsymbol{r'_{j}}$. The exact physical symmetries considered in this analysis are outlined in \Sec{sec:jet_augs}, but for this example, let the augmentation be a random rotation of the dijet event about the beam axis. These two events represent the same underlying physics, as we expect physical events to be symmetric about the beam axis. Hence, $\boldsymbol{r_{j}}$ and $\boldsymbol{r'_{j}}$ are often called \textit{positive pairs}. Therefore we would want the transformer-encoder network to map the event and its augmented version into similar regions of the latent space. In contrast, we would expect the transformer to map the jet event $\boldsymbol{r_{j}}$ and a different jet event $\boldsymbol{r_{k}}$ into different points of the latent space, since we do not expect a high degree of similarity between two arbitrary events. Therefore $\boldsymbol{r_{j}}$ and $\boldsymbol{r_{k}}$ are often called \textit{negative pairs}.
These positive and negative pairs are exactly the pseudo-labels for the contrastive learning method.

These requirements on the transformer-encoder mapping motivate the expression for the contrastive loss:

\begin{align}
\begin{split}
   &\mathcal{L}(\boldsymbol{r_{j}},\boldsymbol{r'_{j}},\boldsymbol{r_{k}},\boldsymbol{r'_{k}},\tau) =
   \\  &-\log{ \left ({\frac{\exp(\textrm{sim}(\boldsymbol{r_{j}},\boldsymbol{r'_{j}}))}{\sum_{\boldsymbol{j}\neq\boldsymbol{k}}[\exp(\textrm{sim}(\boldsymbol{r_{j}},\boldsymbol{r_{k}}))+\exp(\textrm{sim}(\boldsymbol{r_{j}},\boldsymbol{r'_{k}}))]}} \right )}\,.
\label{eq:contrastiv_loss}
\end{split}
\end{align}

We can interpret this loss function as follows: $\textrm{sim}(\boldsymbol{r_{j}},\boldsymbol{r'_{j}})$ calculates the similarity between two latent space representations, where

\begin{align}
\begin{split}
\textrm{sim}(\boldsymbol{r_1}, \boldsymbol{r_2}) = \frac{ \boldsymbol{r_1}\cdot \boldsymbol{r_2} }{ \tau \lVert\boldsymbol{r_1}\rVert \lVert\boldsymbol{r_2}\rVert }.
\end{split}
\end{align}

\noindent The similarity is parameterized by a temperature $\tau$, which balances the numerator and denominator in the contrastive loss. The numerator of the contrastive loss optimizes for \textit{alignment}, which tries to map jets and their augmented versions to similar regions in the latent space. The denominator of the contrastive loss maximizes the \textit{uniformity}, which tries to use up the entirety of the latent space when creating representations (see \Fig{fig:latent_schematic}).

\subsection{\label{sec:jet_augs}Event Augmentations}

We now outline the list of symmetry augmentations used to create physically equivalent latent space jets. 

We define the following single-jet augmentations:

\begin{enumerate}

  \item{\textit{Rotation}}: each jet is randomly (and independently) rotated about its central axis in the $\eta-\phi$ plane.  This is not an exact symmmetry, but correlations between the radiation patterns of the two jets are negligible.
  
  \item{\textit{Distortion}}: each jet constituent is randomly shifted in the $\eta-\phi$ plane. The shift is drawn from a gaussian of mean 0 and standard deviation $\sim 1/p_T$, where $p_T$ is the transverse momentum of the consituent being shifted.  This shift represents the smearing from detector effects.
  
    \item{\textit{Collinear split}}: a small number of the jet constituents (``mothers") are split into two  constituents (``daughters") such that the daughters have $\eta$ and $\phi$ equal to that of the mother, and the transverse momenta of the daughters sum to that of the mother.
            
\end{enumerate}

We define the following event-wide augmentations:

\begin{enumerate}
  
  \item{$\eta$\textit{-shift}}: the dijet event is shifted in a random $\eta$ direction.

  \item{$\phi$\textit{-shift}}: the dijet event is shifted in a random $\phi$ direction.

\end{enumerate}

\begin{figure}
    \centering
    \includegraphics[width=.95\linewidth]{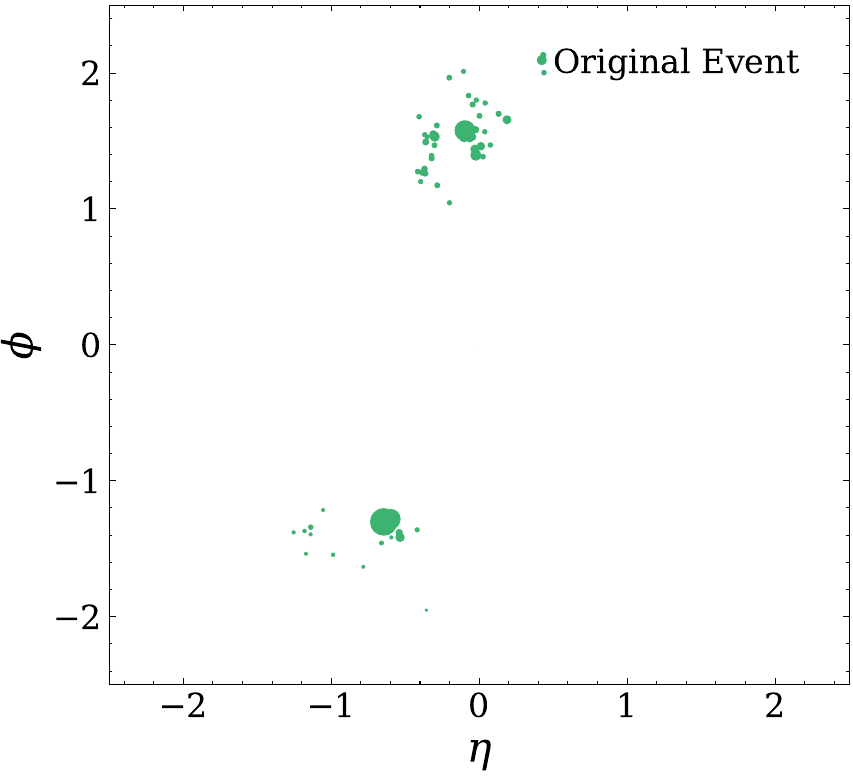}\\
    \includegraphics[width=.95\linewidth]{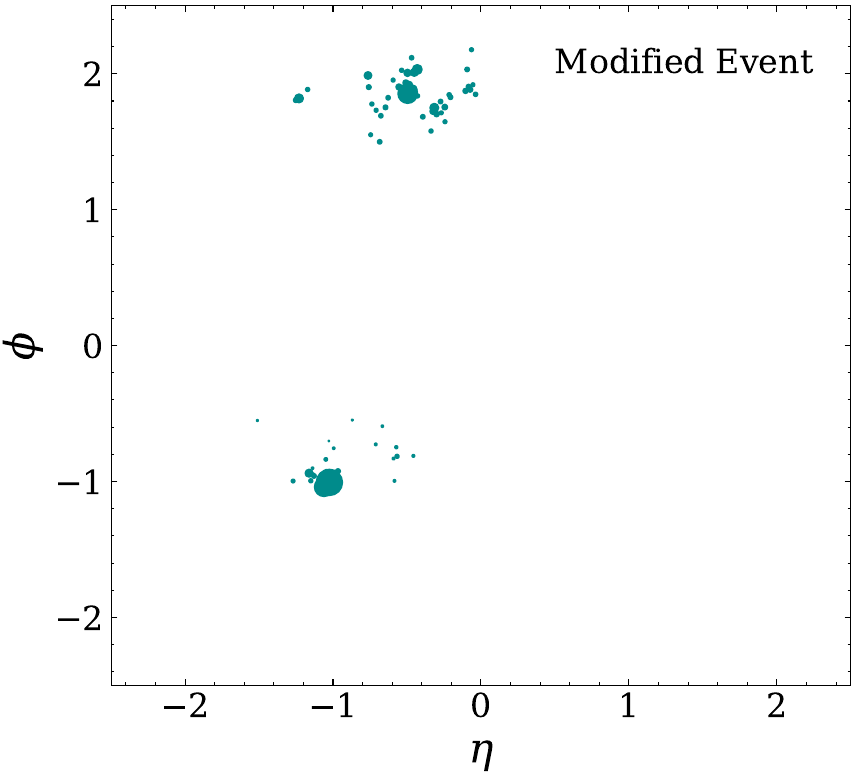}
    \caption{A dijet event before (top) and after (bottom) receiving the set of single-jet and event-wide augmentations. Note that the upper and lower jets have visibly been rotated about their central axes, and the full event has been shifted in the upper-right direction of the $\eta-\phi$ plane.}
    \label{fig:phase_plane_augs}
\end{figure}

Augmentations are applied to each training batch of the transformer. Each jet in the dijet event receives all three of the single-jet augmentaions. The full event then receives both event-wide augmentations. See \Fig{fig:phase_plane_augs} for a visualization of the jet augmentations in the $\eta-\phi$ plane.

\begin{figure}[h]
 \begin{subfigure}[t]{\linewidth}
  \centering\includegraphics[width=\SingleImageWidth\linewidth]{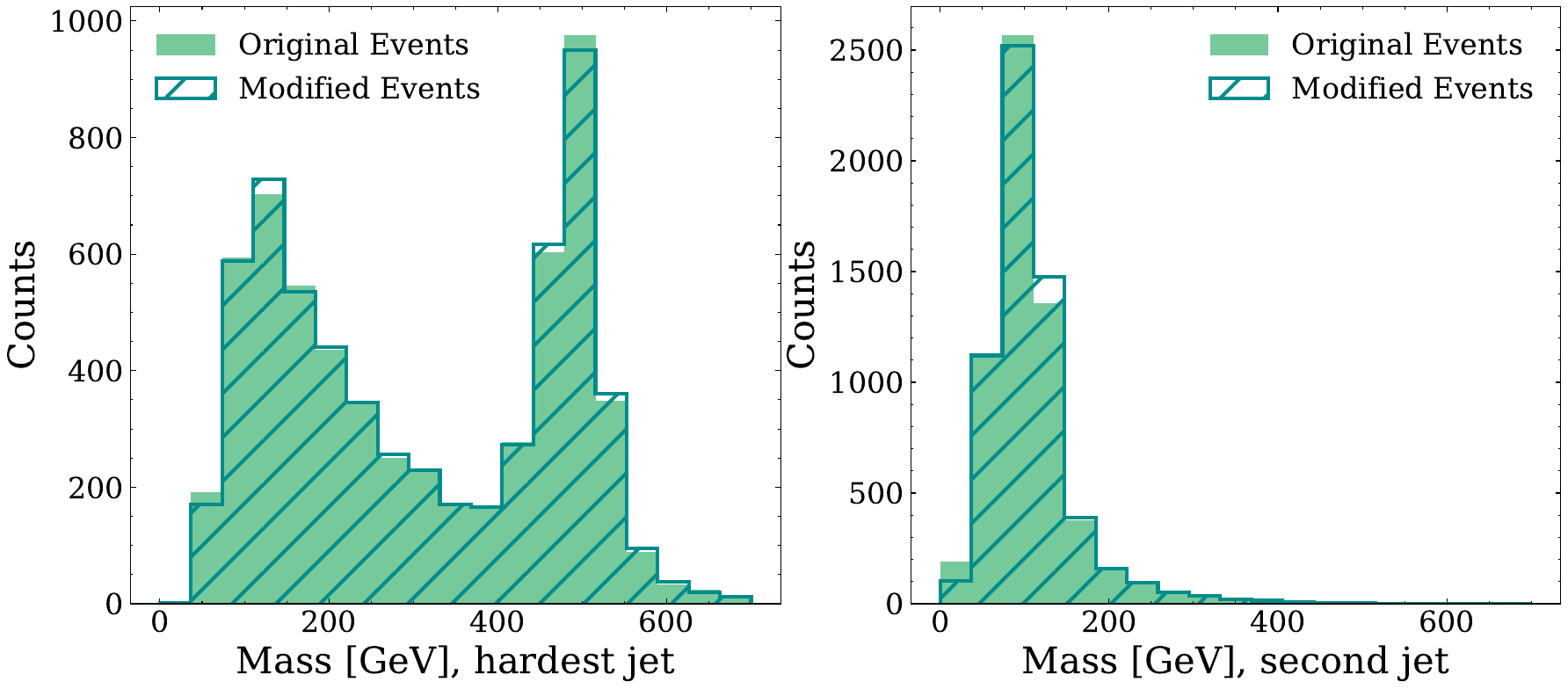}
\caption{Jet mass distributions. }
  \label{fig:jet_masses}
  \end{subfigure}
  \begin{subfigure}[t]{\linewidth}
    \centering\includegraphics[width=\SingleImageWidth\linewidth]{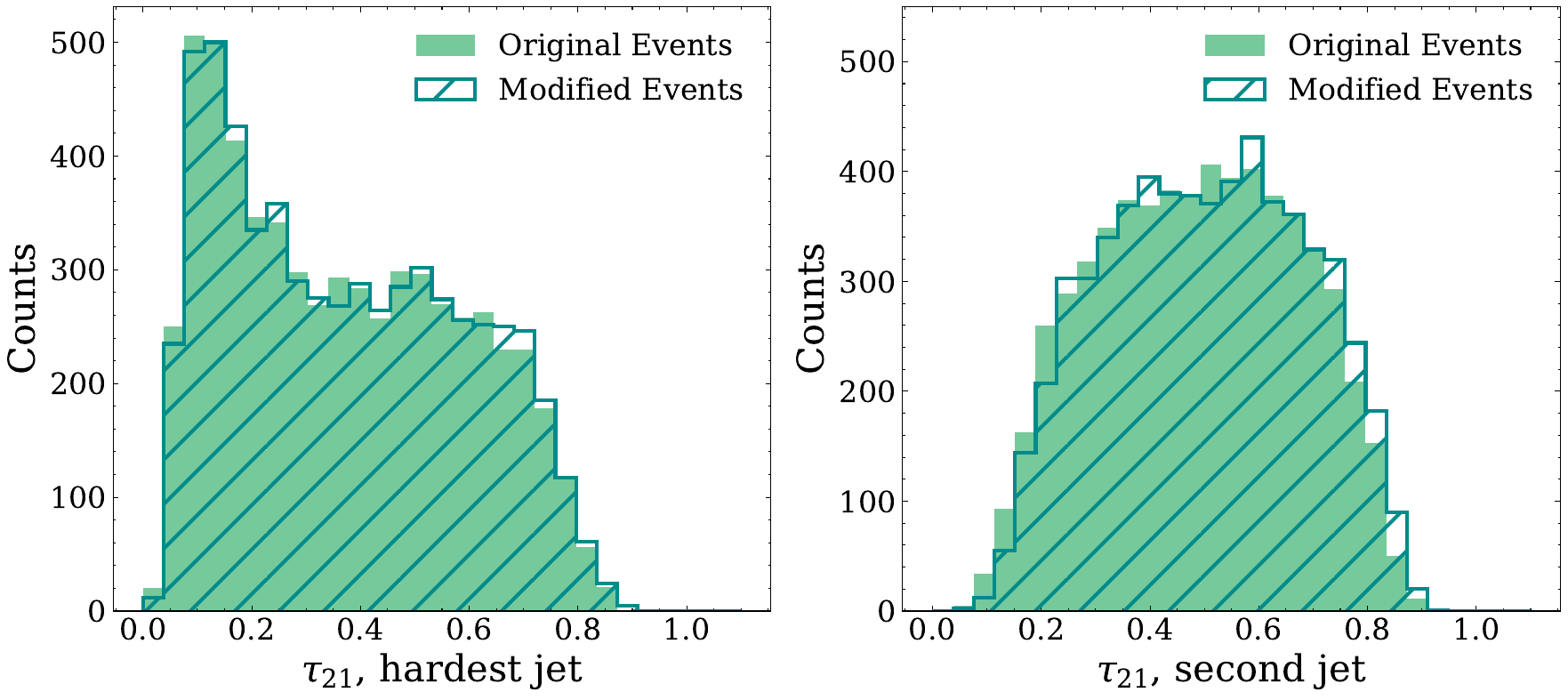}
    \caption{$\tau_{21}$ distributions}
    \label{fig:tau_21}
  \end{subfigure}
  \begin{subfigure}[t]{\linewidth}
    \centering\includegraphics[width=\SingleImageWidth\linewidth]{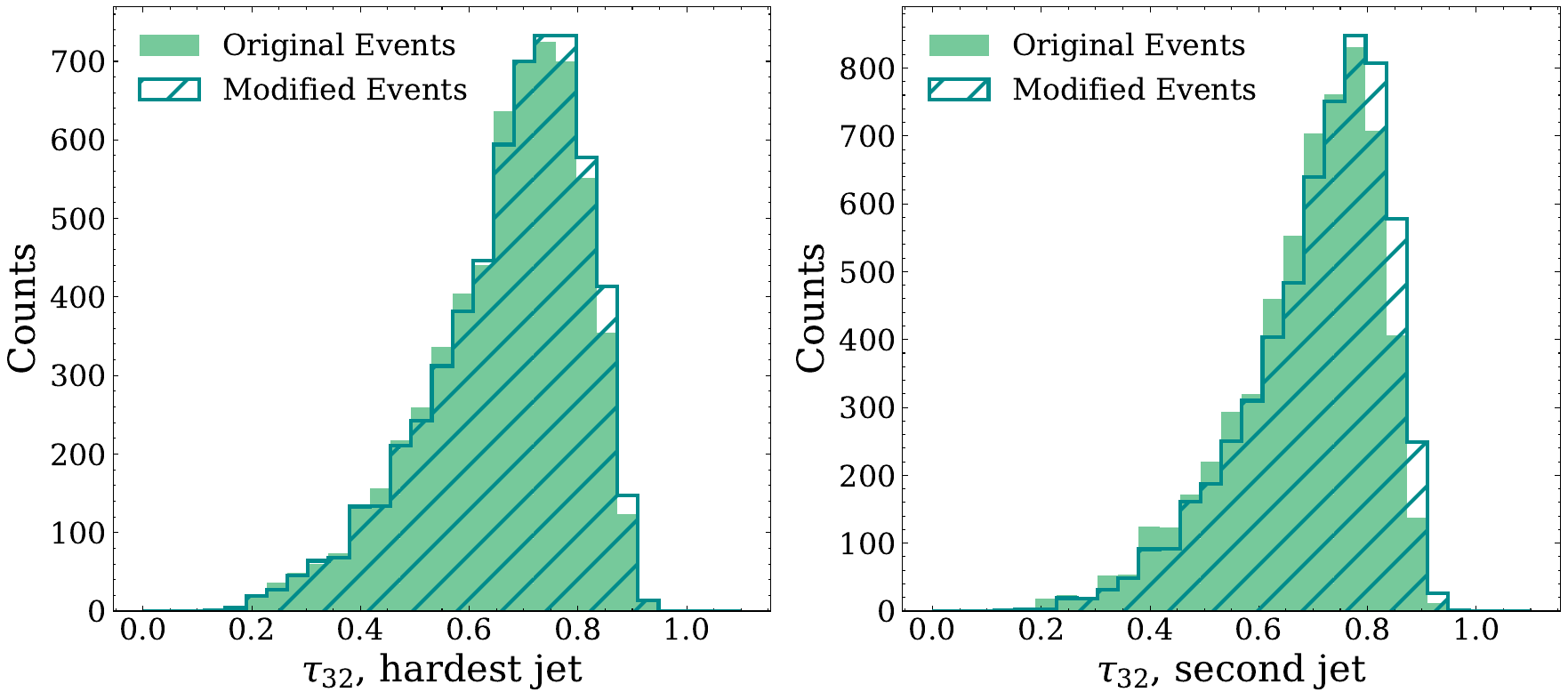}
    \caption{$\tau_{32}$ distributions}
    \label{fig:tau_32}
  \end{subfigure}
  
  \caption{Jet observable distributions for a sample of dijet events, before and after receiving the symmetry augmentations. }

\end{figure}

The jet augmentations are meant to not modify any of the important physical properties of the jets. As a test, we plot the jet masses of the hardest and second hardest jets from a subset of the LHC Olympics dataset in \Fig{fig:jet_masses}, as well as the nsubjettiness variables $\tau_{21}$ and $\tau_{32}$ in \Fig{fig:tau_21} and \Fig{fig:tau_32}, both before and after receiving the jet augmentations and find no significant change in the distributions.

\begin{figure}[h]
  \centering
   \includegraphics[width=\DoubleImageWidth\linewidth]{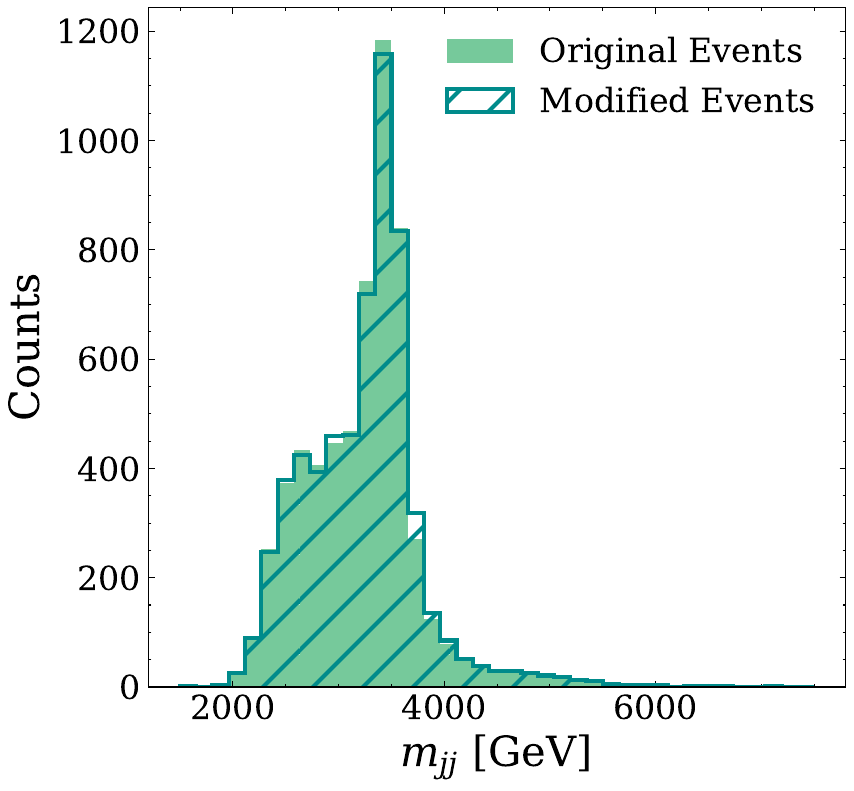}
\caption{Dijet mass ($m_{JJ}$) distributions for a sample of dijet events, before and after receiving the symmetry augmentations.}
  \label{fig:mjj}

\end{figure}

We also plot $m_{JJ}$ for the dijet system in \Fig{fig:mjj}, again finding good agreement before and after the augmentations are applied. This confirms that our set of jet augmentations can be seen as true symmetry transformations of the dijet events.

\subsection{\label{sec:training}Training procedure}

We train the transformer-encoder network on a dataset of 50,000 background dijet events and up to 50,000 signal events, optimized on the contrastive loss in Eq. \ref{eq:contrastiv_loss}. The batch size is set to 400, which is the largest possible given the computing resources available. The network is trained with a learning rate of 0.0001, an early stopping parameter of 20 epochs, and a temperature parameter $\tau$ of 0.1. All of these hyperparameters were empirically found to deliver the best transformer performance.  The transformer-encoder network is implemented using \texttt{Pytorch} 1.10.0 \cite{NEURIPS2019_9015} and optimized with \texttt{Adam} \cite{adam}. Jet augmentations are applied batchwise, with each dijet event receiving a different randomized augmentation. 

We also construct a binary classification dataset used to evaluate the latent space jet representations. This dataset consists of 85,000 signal and background dijet events each. 
We consider two types of binary classification tasks. Fully connected binary classifiers (FCN's) are implemented in \texttt{Pytorch} and optimized with \texttt{Adam}. These networks consist of three linear layers of sizes (64, 32, 1) with $\textsc{ReLu}$ activation, a dropout of 0.1 between each layer, and a final sigmoid layer. The FCN is trained with a batch size of 400, a learning rate of 0.001, and an early stopping parameter of 5 epochs. Linear classifier tests (LCT's) are implemented in \texttt{scikit-learn} \cite{scikit}. Both binary classification tasks discriminate signal from background in the latent space and have access to signal and background labels (i.e. are fully supervised).

We further define a ``standard test" dataset consisting of 10,000 signal and background events each. There is no overlap of events in the standard test dataset with those in the transformer training or binary classifier training sets.

\subsection{\label{sec:anomaly_detection}Anomaly Detection}

The usefulness of the latent space dijet representations is evaluated in a realistic model agnostic anomaly detection search setup. \textsc{CWoLa} (Classification Without Labels) \cite{Metodiev:2017vrx} is a weakly supervised training method that allows for signal versus background discrimination in cases where training samples of pure signal and background cannot be provided. Such a scenario might occur in resonant bump-hunting, where it is common to define signal and sideband regions, both of which will have a non-negligible fraction of background events~\cite{Collins:2018epr,Collins:2019jip}. The authors of \Ref{Metodiev:2017vrx} show that a classifier that is trained on two mixed samples (each with a different signal fraction) is in fact maximally discriminatory for classifying signal from background.

In \Sec{sec:cwola}, we run a \textsc{CWoLa} training procedure on the latent space representations. Our mixtures consist of one background-only sample and one sample with a suppressed signal fraction representing a mixture of background and a rare unknown anomaly.  This is the ideal anomaly detection setup in which a sample of pure background can be generated (and is also the starting point of Refs~\cite{DAgnolo:2018cun,DAgnolo:2019vbw,dAgnolo:2021aun}).  In practice, this may not be possible, and other methods must be used to obtain this dataset directly from data using sideband information~\cite{Nachman:2020lpy,Andreassen:2020nkr,Stein:2020rou,1815227,Hallin:2021wme,Raine:2022hht}.

\section{\label{sec:latent_space}Evaluating the latent space representations}

We evaluate the ability of our transformer-encoder network to faithfully translate dijet events into a latent space in the following way: we first train the network to embed event space particle collisions into a reduced-dimension latent space. We then perform a binary classification task on the latent space. We test the sensitivity of this setup to the amount of signal present in the training of the transformer as well as in the training of the classifier.  The latter test demonstrates the anomaly detection capability of the approach.

\subsection{\label{}Quantifying the effect of each augmentation}

\begin{figure}
    \centering
    \includegraphics[width=\SingleImageWidth\linewidth]{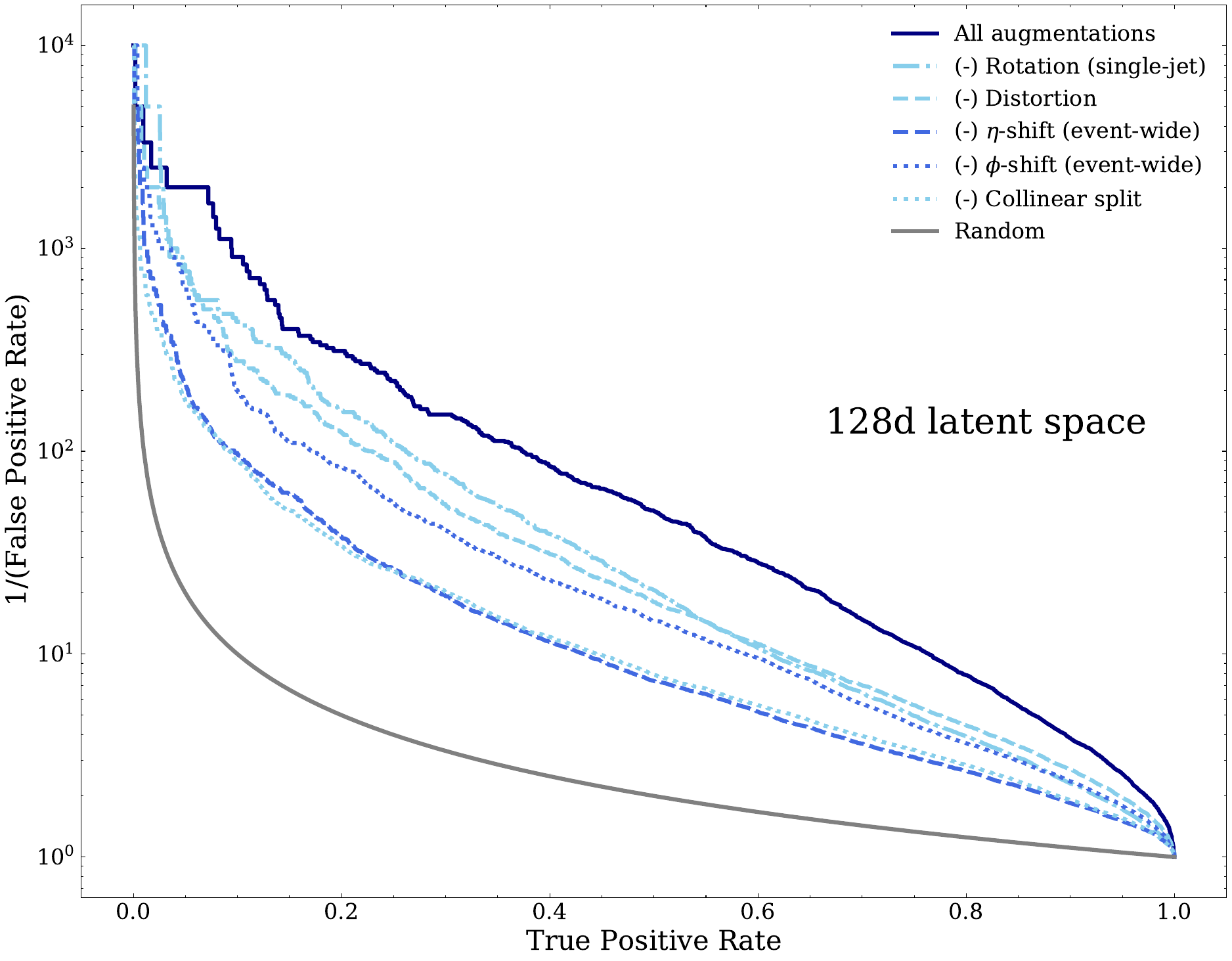}
    \caption{Classifier efficiency curves for a fully connected binary classifier (FCN) run on latent space dijet representations trained with all except the indicated augmentation. All of the five augmentations appear to contribute significantly to the transformer-encoder network performance.}
    \label{fig:subtraction_study_128_NN}
\end{figure}

As a first study, we explore the importance of each of the five symmetry augmentations outlined in \Sec{sec:jet_augs}. In \Fig{fig:subtraction_study_128_NN}, we plot the rejection (1 / false positive rate)
versus the true positive rate for a FCN trained on the latent space dijet representations. Each curve in the figure represents a transformer-encoder network trained with all of the symmetry augmentations \textit{except} the one indicated. The transformer-encoder network is trained on 50,000 signal and 50,000 background dijet events, and the dimension of the transformer latent space is held at 128 dimensions.

In general, the performance of the transformer-encoder (as quantified by the receiver operating characteristic (ROC) area under the curve (AUC) for the FCN) drops sizably if any of the symmetry augmentations is not used during the transformer-encoder network training. The worst-performing transformers are those that do not receive the event-wide $\eta$-shift or collinear split augmentations. However, the decrease in AUC is significant for every removed augmentation. It is likely that the addition of symmetry augmentations would lead to further improvement in the transformer performance.

The AUC scores for the FCN's trained on latent space representations are given in \Tab{tab:AUC_scores_subtractive}. We additionally include as a performance metric the maximum of the significance improvement characteristic (\textrm{max(SIC)}), defined as $\max(\frac{\textrm{true positive rate}}{\sqrt{\textrm{false positive rate}}})$. The max(SIC) can be seen as the multiplicative factor by which signal significance improves after performing a well-motivated cut on the dataset.

\begin{table}
    \centering
    \begin{tabular}{|l|r|r|}
        \hline 
        Removed Augmentation & AUC & max(SIC) \\
        \hline 
        \hline 
        (None) & 0.918 & 3.805 \\
        Distortion & 0.872 & 2.376 \\
        Rotation (single-jet) & 0.860 & 2.677 \\
        $\phi$-shift (event-wide) & 0.851 & 1.949 \\
        Collinear split & 0.800 & 1.430 \\
        $\eta$-shift (event-wide) & 0.791 & 1.387 \\
        
        \hline 
        
    \end{tabular}
    \caption{ROC AUC and max(SIC) scores for a binary classifier trained to discriminate signal from background on the LHC Olympics dataset. The performance scores decrease sizably if any of the five symmetry augmentations are not used when training the transformer-encoder network.}
    \label{tab:AUC_scores_subtractive}
\end{table}

\subsection{\label{}Exploring the dimensionality of the latent space}

We next gauge how the size of the latent space affects the usefulness of the representations. The latent space jet representations would ideally be lower in dimension than the physical space versions so as to save computational processing time by removing nonessential degrees of freedom from a given dataset. However, a latent space embedding with too few dimensions might not contain enough parameters to encode the essential physical dynamics of the jets.

\begin{figure}
    \centering
    \includegraphics[width=\SingleImageWidth\linewidth]{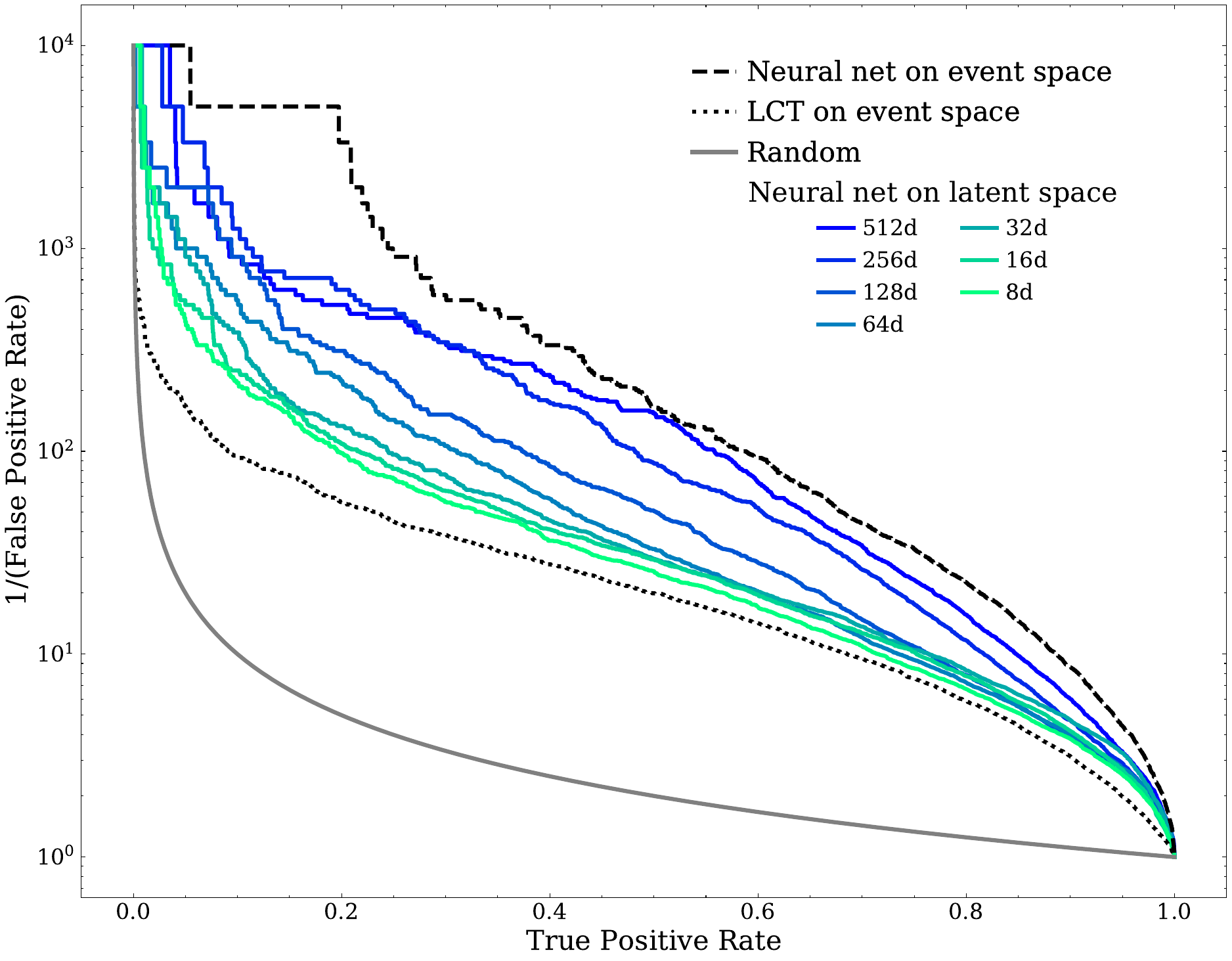}
    \caption{Classifier efficiency curves for a FCN run on the latent space dijet representations. The performance of the binary classifier increases with the dimension of the latent space. For comparison, we also provide efficiency curves for a FCN and a linear classifier test (LCT) run on the event space dijets.}
    \label{fig:fpr_tpr_dim_scan}
\end{figure}

In \Fig{fig:fpr_tpr_dim_scan}, we plot the rejection versus the true positive rate for a FCN trained on the latent space dijet representations. (See \Fig{fig:fpr_tpr_dim_scan_LCT} in \App{sec:latent_space_lct} for curves from a linear classifier.) We scan the latent space size in powers of two from 512 down to 8 dimensions. For all latent space dimensions, the transformer-encoder network is trained on 50,000 signal and 50,000 background dijet events.  

The performance of the transformer-encoder improves as the dimension of the latent space increases. We find that a FCN trained on latent space jet representations cannot outperform a FCN trained on event space jet representations, but it can outperform a LCT trained on event space jet representations.
Perhaps more striking is that the linear classifier trained on the compressed latent representations outperforms the linear classifier trained on the full event space data.  This indicates that the self-supervised representations are highly expressive despite being compressed, and it agrees with the top-tagging results obtained in Ref.~\cite{Dillon:2021gag}.

A selection of AUC scores for the FCN's trained on latent space representations are given in \Tab{tab:AUC_scores_dim_scan}. The table also contains scores for a LCT trained on the latent space representations, as well as a binary classifier constructed from the transformer architecture with an additional sigmoid function as the final layer (Trans+BC), trained using the Binary Cross Entropy loss.
The Trans+BC network has access to all of the input particles and is not trained post-hoc on the self-supervised latent space, so we expect it to perform the best of all configurations.  The hope is that the FCN performance is as close as possible to the performance of the Trans+BC (on the largest latent space).

\Tab{tab:AUC_scores_dim_scan} does show a performance gap between a FCN trained on the latent space and the Trans+BC network. In \App{sec:datahungry}, we evaluate the performance of both such networks when trained on increasing amounts of data. In both cases, we find that the networks are ``data-hungry"; in other words, the classifier performances increase with the amount of training data, and the performances do \textit{not} saturate when trained on the 85,000 signal and 85,000 background dijet events sampled from the LHCO dataset. Therefore the performance of the FCN could likely reach that of the Trans+BC network with a larger training dataset than what was used in this study.

\begin{table}
    \centering
    \begin{tabular}{|c|c|c|c|r|}
        \hline 
        Training set & Training dim. & Classifier & AUC & max(SIC) \\
        \hline 
        \hline 
        \multirow{2}{*}{Particle space} & \multirow{2}{*}{506 $\pm$ 174} & FCN & 0.958 & 15.401 \\
        & & LCT & 0.883 & 2.277 \\
        \hline 
        \multirow{9}{*}{Latent space} & \multirow{3}{*}{8} & FCN & 0.904 & 2.542 \\
        & & LCT & 0.841 & 1.882 \\
        & & Trans+BC & 0.955 & 7.608\\
        \cline{2-5}
        & \multirow{3}{*}{64} & FCN & 0.915 & 3.163 \\
        & & LCT & 0.816 & 1.799 \\
        & & Trans+BC & 0.960 & 7.238 \\
        \cline{2-5}
        & \multirow{3}{*}{512} & FCN & 0.945 & 6.396 \\
        & & LCT & 0.926 & 4.624 \\
        & & Trans+BC & 0.968 & 13.862 \\
        \hline 
        
    \end{tabular}
    \caption{ROC AUC and max(SIC) scores for a binary classifier trained to discriminate signal from background on the LHC Olympics dataset. FCN = Fully Connected (Dense) Neural Network; LCT = Linear Classifier Test; Trans+BC = transformer architecture trained on the Binary Cross Entropy loss. The particle space training dimension is (avg. no. of nonzero entries) $\pm$ (std. dev. of nonzero entries) per LHC Olympics event.}
    \label{tab:AUC_scores_dim_scan}
\end{table}

\subsection{\label{}Varying the amount of training signal}

In practice, we want to use the transformer-encoder network for model-agnostic anomaly detection. In this case, we would not be able to train the transformer on a known signal fraction, as the training data would contain an unknown (and extremely tiny, if any) percentage of BSM signal. 
It is therefore useful to see if the transformer-encoder network is effective at translating rare events into a latent space.  
We might expect this to be true if the transformer is learning only generic features about collider events that hold for both signal and background events.  This is encouraged by the universality of the symmetry augmentations in the contrastive loss.

\begin{figure}
  \centering
  \begin{subfigure}{.95\linewidth}
    \includegraphics[width=\SingleImageWidth\columnwidth]{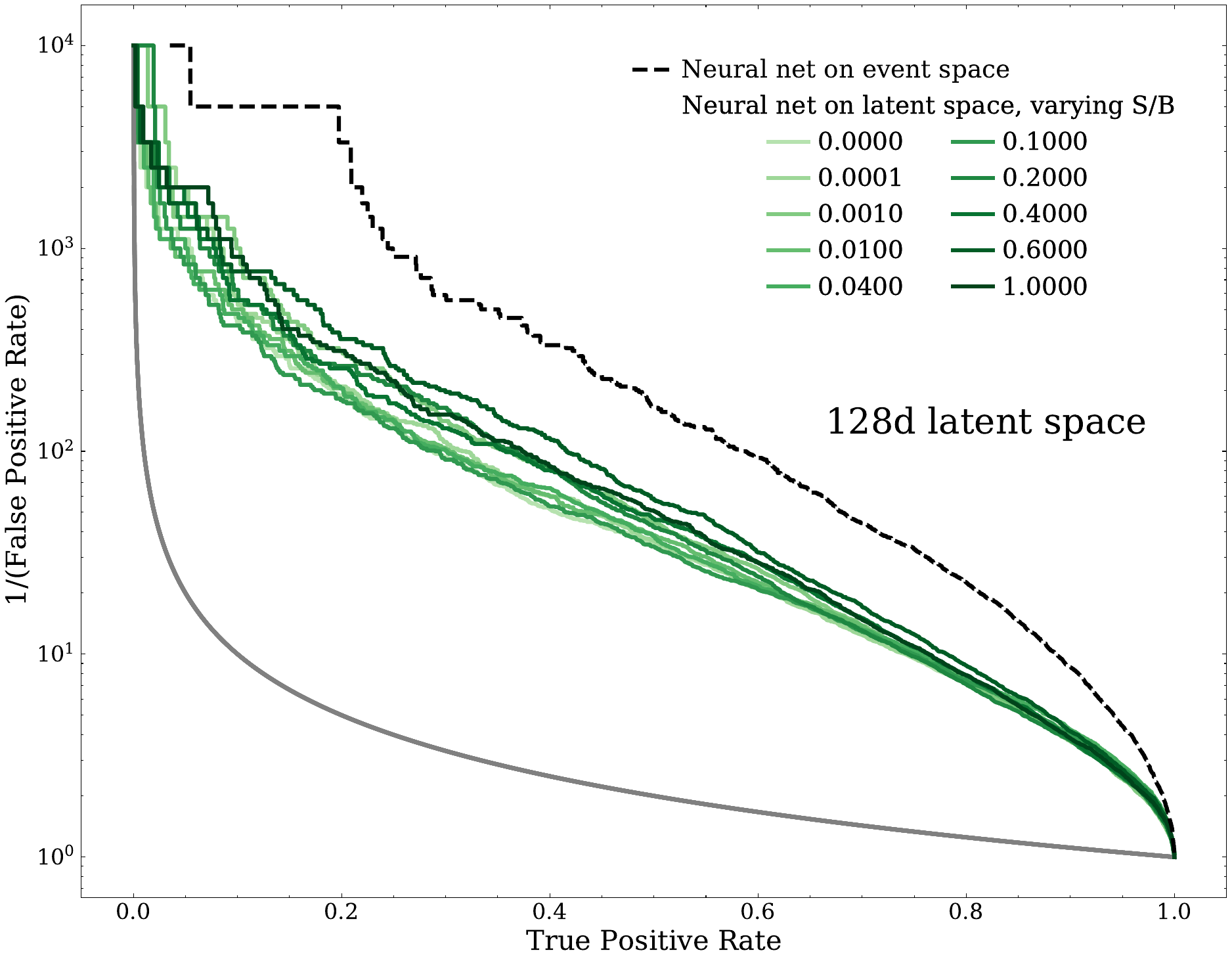}
    \caption{128d latent space}
    \label{fig:SvB_128}
  \end{subfigure}\\
  \begin{subfigure}{.95\linewidth}
    \includegraphics[width=\SingleImageWidth\columnwidth]{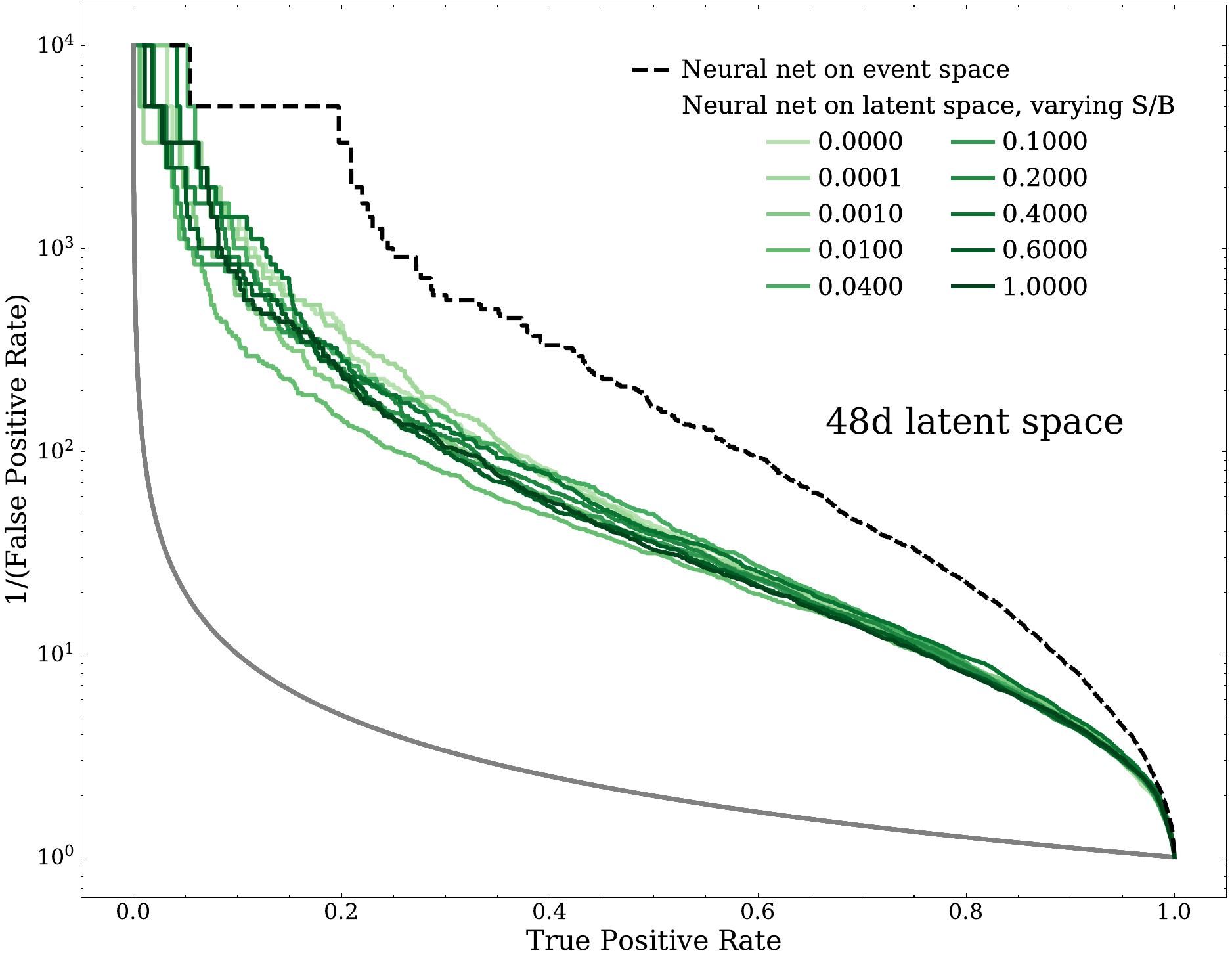}
    \caption{48d latent space}
    \label{fig:SvB_48}
  \end{subfigure}
  \caption{Classifier efficiency curves for a latent space FCN classifier trained on varying amounts of signal fraction. The classifier performance is robust with respect to the signal fraction of the training data. This implies that the transformer-encoder network can be trained on background alone.}
  \label{fig:SvB}
\end{figure}

In \Fig{fig:SvB_128}, we hold the dimension of the transformer latent space fixed at 128, then scan the signal to background ratio S/B down from 1.0 to 0.0. In \Fig{fig:SvB_48}, we repeat the previous steps for a transformer latent space of dimension 48.  (See \Fig{fig:SvB_LCT} in \App{sec:latent_space_lct} for curves from a linear classifier.) Note that for this study, the transformer-encoder network is always trained on 50,000 background dijet events, but the number of signal dijet events changes with the signal S/B ratio. We find that the classifier performance is robust with respect to the signal to background ratio, as was found in \Ref{Dillon:2021gag}. This demonstrates that the transformer-encoder network can be trained on background alone and still faithfully model rare signal events.

\section{\label{sec:cwola}Anomaly detection}

We now test the usefulness of the latent space jet representations in a more practical setting by performing a \textsc{CWoLa}-style anomaly search. To create the latent space representations, we use a transformer-encoder network trained on a background-only sample of 50,000 dijet events. As before, we use the ``standard test" dataset of 10,000 signal and 10,000 background events for all binary classifier tests.

We first create a baseline against which to compare the \textsc{CWoLa} analyses by carrying out a self-supervised binary classification task in the event space representation. For this study, we use 42,500 signal and 42,500 background dijet events.

\begin{figure*}[h]
  \centering

  \begin{subfigure}[t]{.7\linewidth}
    \centering\includegraphics[width=\SingleImageWidth\linewidth]{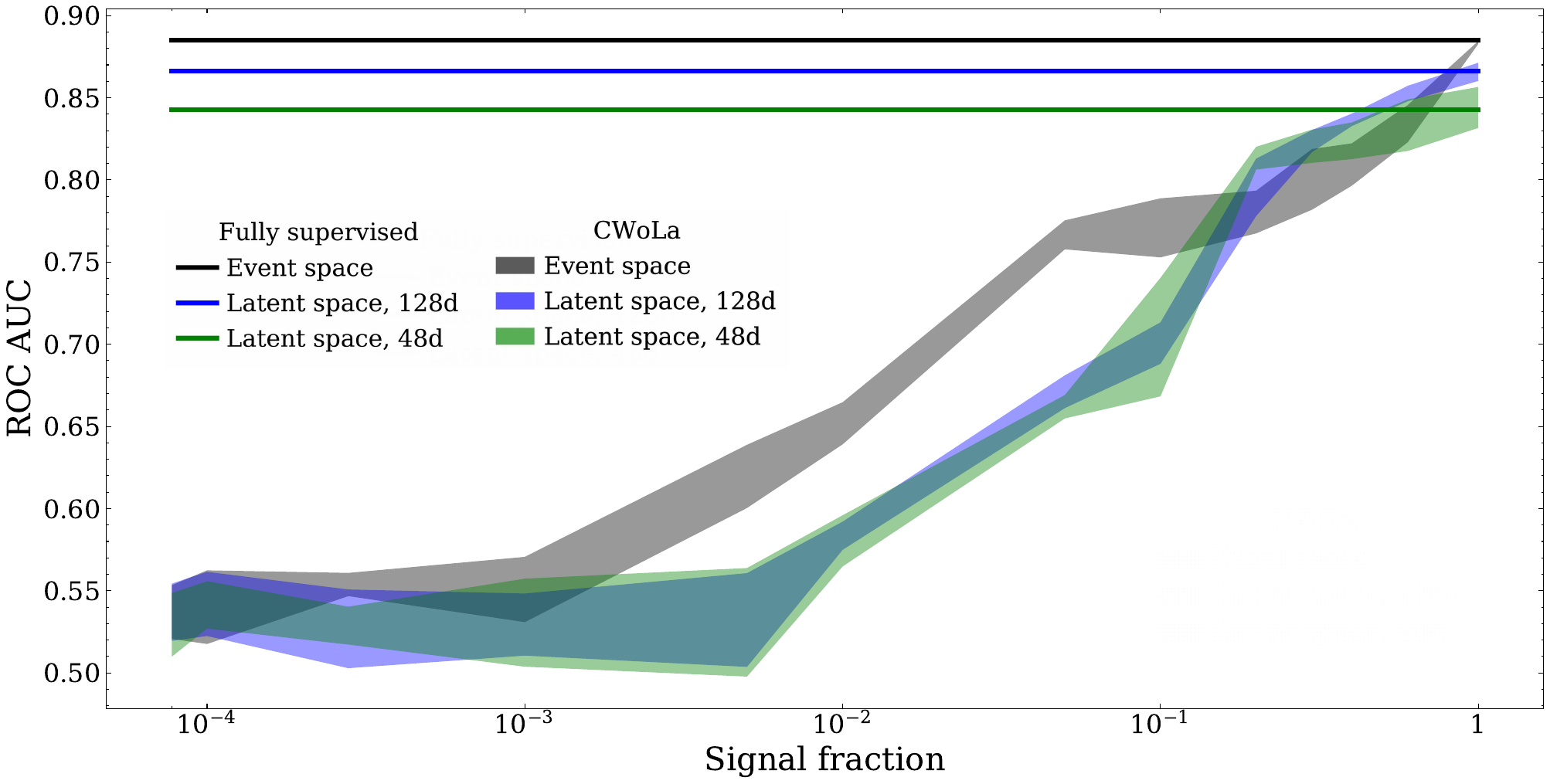}
    \caption{ROC AUC training metric.}
     \label{fig:JetCLR_CWoLa_ROC}
  \end{subfigure}
  \\
    \begin{subfigure}[t]{.7\linewidth}
    \centering\includegraphics[width=\SingleImageWidth\linewidth]{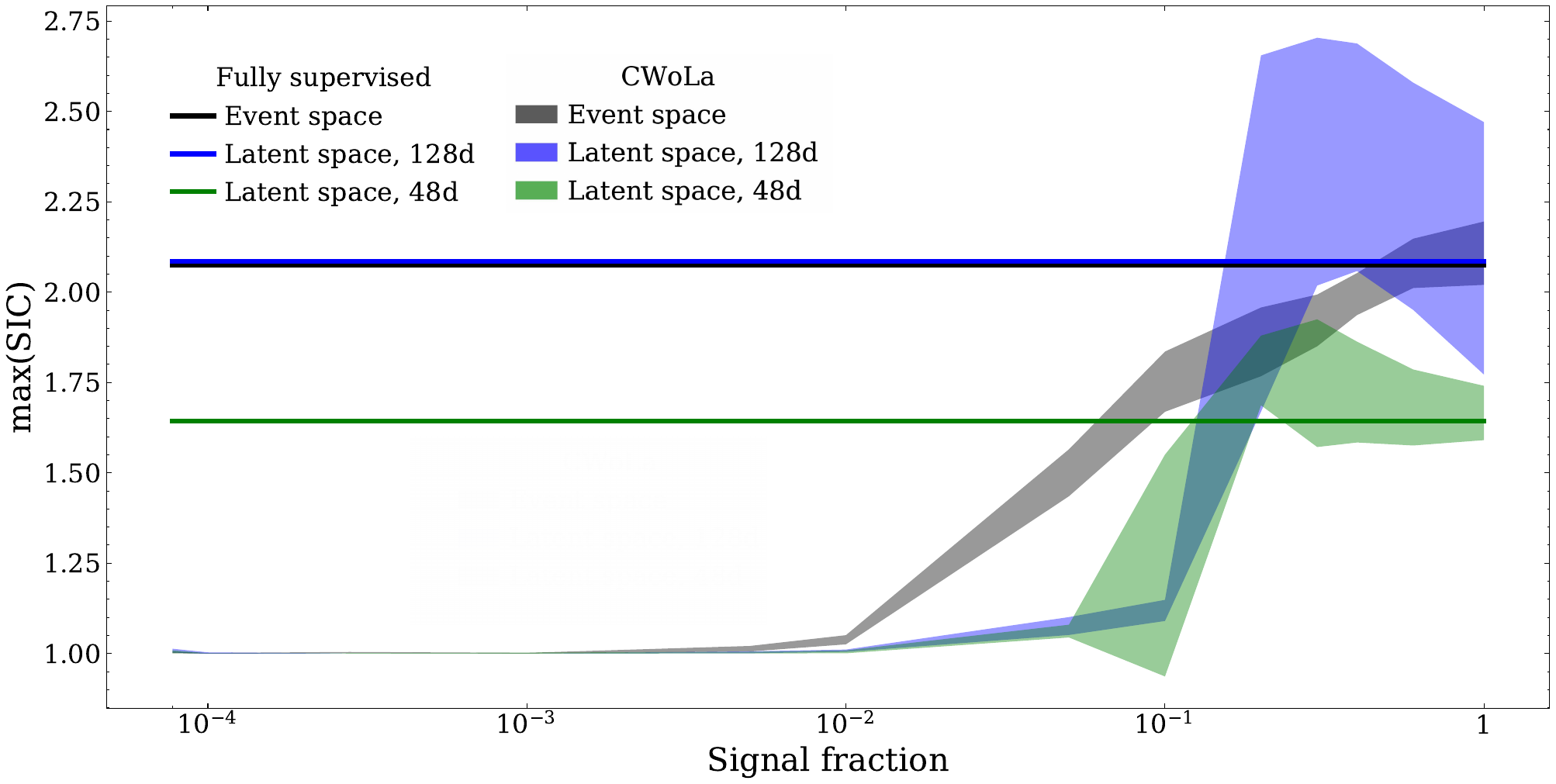}
    \caption{\textrm{max(SIC)} training metric.}
    \label{fig:JetCLR_CWoLa_maxsic}
  \end{subfigure}
  \\
   \begin{subfigure}[t]{.7\linewidth}
    \centering\includegraphics[width=\SingleImageWidth\linewidth]{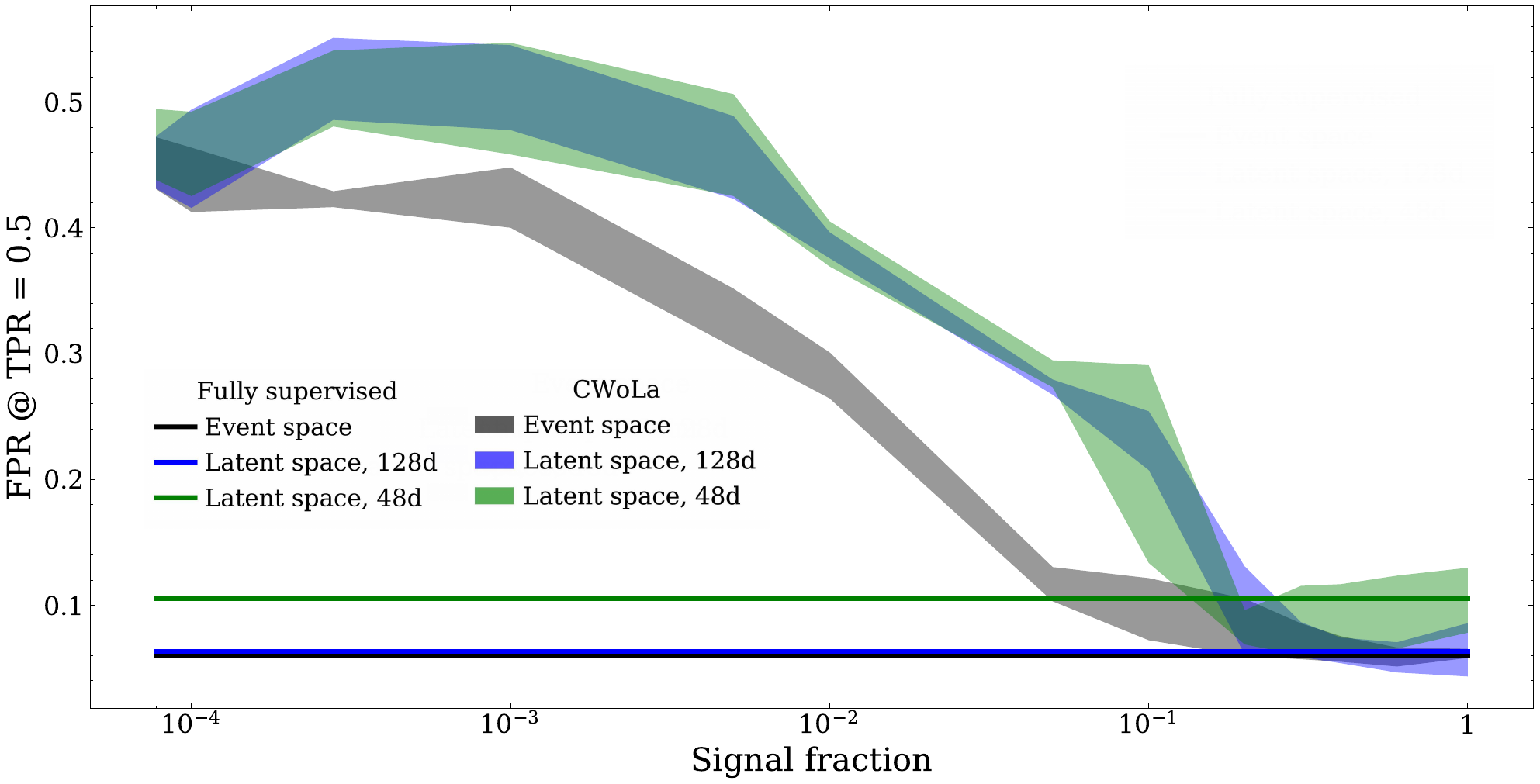}
    \caption{False positive rate at a true positive rate of 50\%.}
     \label{fig:JetCLR_CWoLa_FPR}
  \end{subfigure}

  \caption{Various metrics for evaluating a \textsc{CWoLa} weakly-supervised classifier trained to discriminate a background-only sample from a signal+background sample with a variable signal fraction. The classifier run on the event space representations slightly outperforms one run on the latent space representations, especially at low signal fractions.}
  \label{fig:cwola}

\end{figure*}

For the anomaly-detection analysis, we set one \textsc{CWoLa} ``mixed sample" to be a set of 42,500 background-only dijets (the same as in the self-supervised task). The other mixed sample is a mixture of 42,500 signal and background dijets, with the signal fraction scanned from 0\% to 100\%. We run the analysis three times, once for the event space dijets and once each for latent space dijets at 128 and 48 dimensions.  The evaluation of the performance is always computed with pure signal and background labels.  Comparisons of the \textsc{CWoLa} classifier performances are shown in \Fig{fig:cwola}. In \Fig{fig:JetCLR_CWoLa_ROC}, we use the ROC AUC as a metric for evaluating the \textsc{CWoLa} classifier; in \Fig{fig:JetCLR_CWoLa_maxsic}, we use instead use the \textrm{max(SIC)}as the metric; in \Fig{fig:JetCLR_CWoLa_FPR}, we provide the false positive rate at a fixed true positive rate of 50\%.  


We find that the \textsc{CWoLa} weakly supervised classifier performance of a small dimensional latent space is comparable to (but cannot match) that of the full particle event space, with an improvement in performance for a larger dimension latent space as evaluated by the \textrm{max(SIC)} metric. The most notable difference between the classifier performance on latent space vs. event space is that in the former case, the classifier performance diminishes to no better than random at a higher signal fraction for the training data. (This is indicated by the ROC AUC dropping to 0.5, the max(SIC) dropping to 1, and the FPR @ TPR = 0.5 dropping to 0.5.) More specifically, the small-dimension latent space classifier hits random performance at a signal fraction of just below $10^{-2}$, while the event space classifier does better than random at all nonzero signal fractions.

Overall, the classifier performances at anomaly-level signal fractions as shown in \Fig{fig:cwola} are lower than what has been seen in other recent anomaly-detection methods on the LHCO data. In fact, the SIC curves shown in Refs.~\cite{Nachman:2020lpy,Andreassen:2020nkr,Stein:2020rou,1815227,Hallin:2021wme,Raine:2022hht} are typically an order of magnitude greater than those in \Fig{fig:cwola}. However, such curves were constructed by training on standard jet observables (e.g. $m_J$, $\tau_{12}$), and thus represent training methods that are inherently model-dependent. Evidently, this performance decrease from model-dependent anomaly detection methods is the price to pay on this particular signal for using a more widely-applicable, model-agnostic method.

There exist a number of avenues for future work to improve on this contrastive-learning trained classifier. For one: we have considered a small set of symmetry augmentations specific to dijet events. However, additional augmentations for dijet events could be added to the contrastive loss. Alternatively, a different selection of augmentations that leads to an even more general event representation could be chosen. As another avenue: we mentioned earlier (and illustrate in \App{sec:datahungry}) that the transformer-encoder network is data-hungry. It would therefore be reasonable to expect an improvement in classifier performance if the training dataset were made larger.

\section{\label{sec:conclusions}Conclusions}

In this paper, we have used transformer-encoder neural networks to embed entire collider events into low and fixed-dimensional latent spaces.  This embedding is constructed using self-supervised learning based on symmetry transformations.  Events that are related by symmetry transformations are grouped together in the latent space while other pairs of events are spread out in the latent space.

We have shown that the latent space preserves the essential properties of the events for distinguishing certain BSM events from the SM background.  This latent space can then be used for a variety of tasks, including anomaly detection.  We have shown that anomalies can still be identified in the reduced representation as long as there is enough signal in the dataset. For the particular signal model studied, the required amount of signal is much higher than reported by other studies using high-level features.  This illustrates the tradeoff between signal sensitivity and model specificity.  Our reduced latent space knows nothing of particular BSM models and is thus broadly useful but not particularly sensitive.  Future work that explores the continuum of approaches by adding more augmentations to the contrastive learning may result in superior performance for particular models in the future.

\section*{Code availability}

\noindent The code can be found at \\ \href{https://github.com/rmastand/JetCLR\_AD}{https://github.com/rmastand/JetCLR\_AD}.

\begin{acknowledgments}

We thank Jernej Kamenik and David Shih for useful feedback on the manuscript.

B.M.D was supported by a Postdoctoral Research Fellowship from the Alexander von Humboldt Foundation.
B.N. and R.M. were supported by the Department of Energy, Office of Science under contract number DE-AC02-05CH11231. This material is based upon work supported by the National Science Foundation Graduate Research Fellowship Program under Grant No. DGE 2146752. Any opinions, findings, and conclusions
or recommendations expressed in this material are those of the authors and do not necessarily reflect the views of the National Science Foundation.

\end{acknowledgments}


\appendix

\section{Evaluating the latent space with a linear classifier test}
\label{sec:latent_space_lct}

\begin{figure}[h!]
    \centering
    \includegraphics[width=\SingleImageWidth\linewidth]{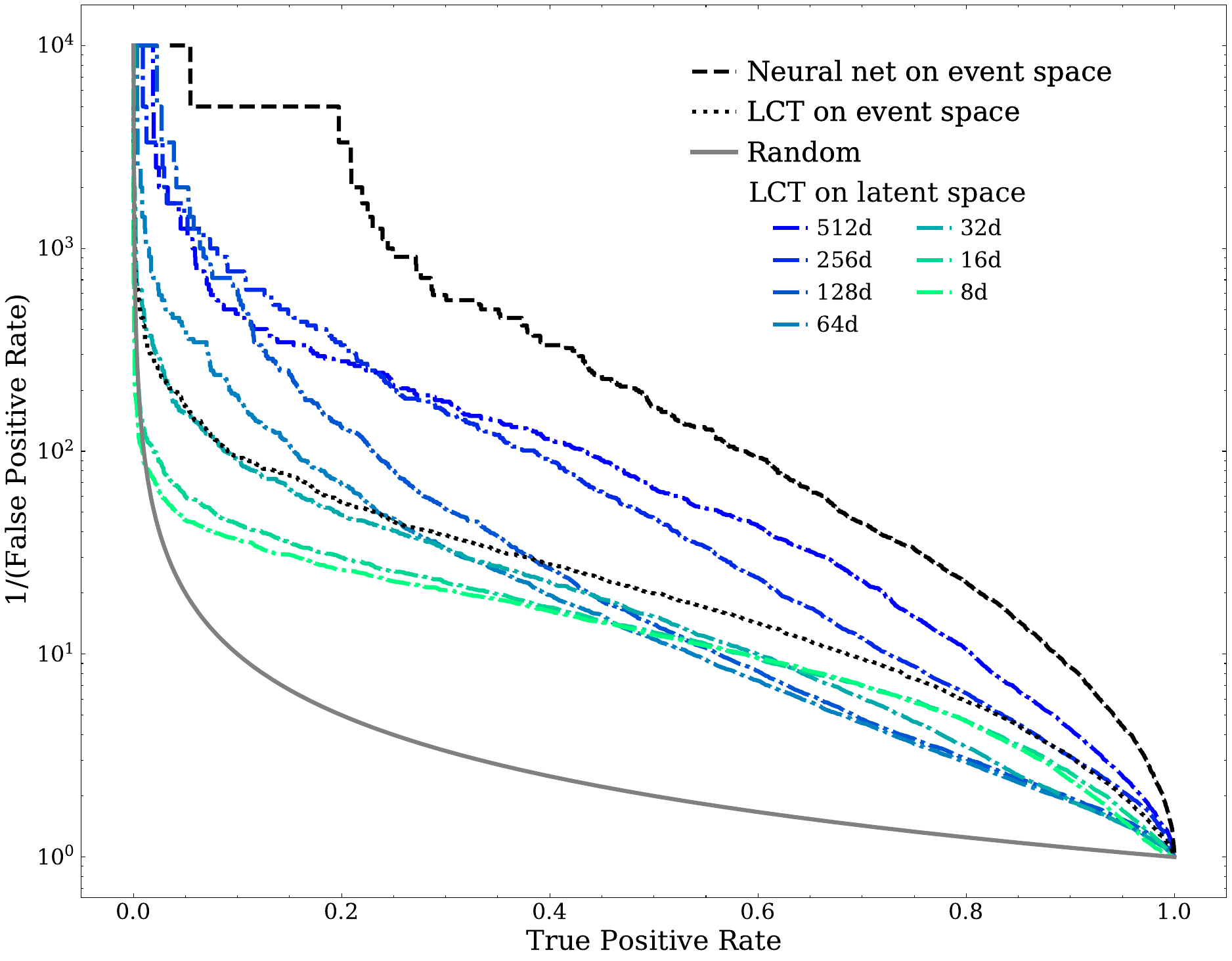}
    \caption{Classifier efficiency curves for a linear classifier test (LCT) run on the latent space dijet representations. For comparison, we also provide efficiency curves for a FCN and a LCT run on the event space dijets.}
    \label{fig:fpr_tpr_dim_scan_LCT}
\end{figure}

\begin{figure*}
  \centering
  \begin{subfigure}{.49\linewidth}
    \includegraphics[width=\SingleImageWidth\columnwidth]{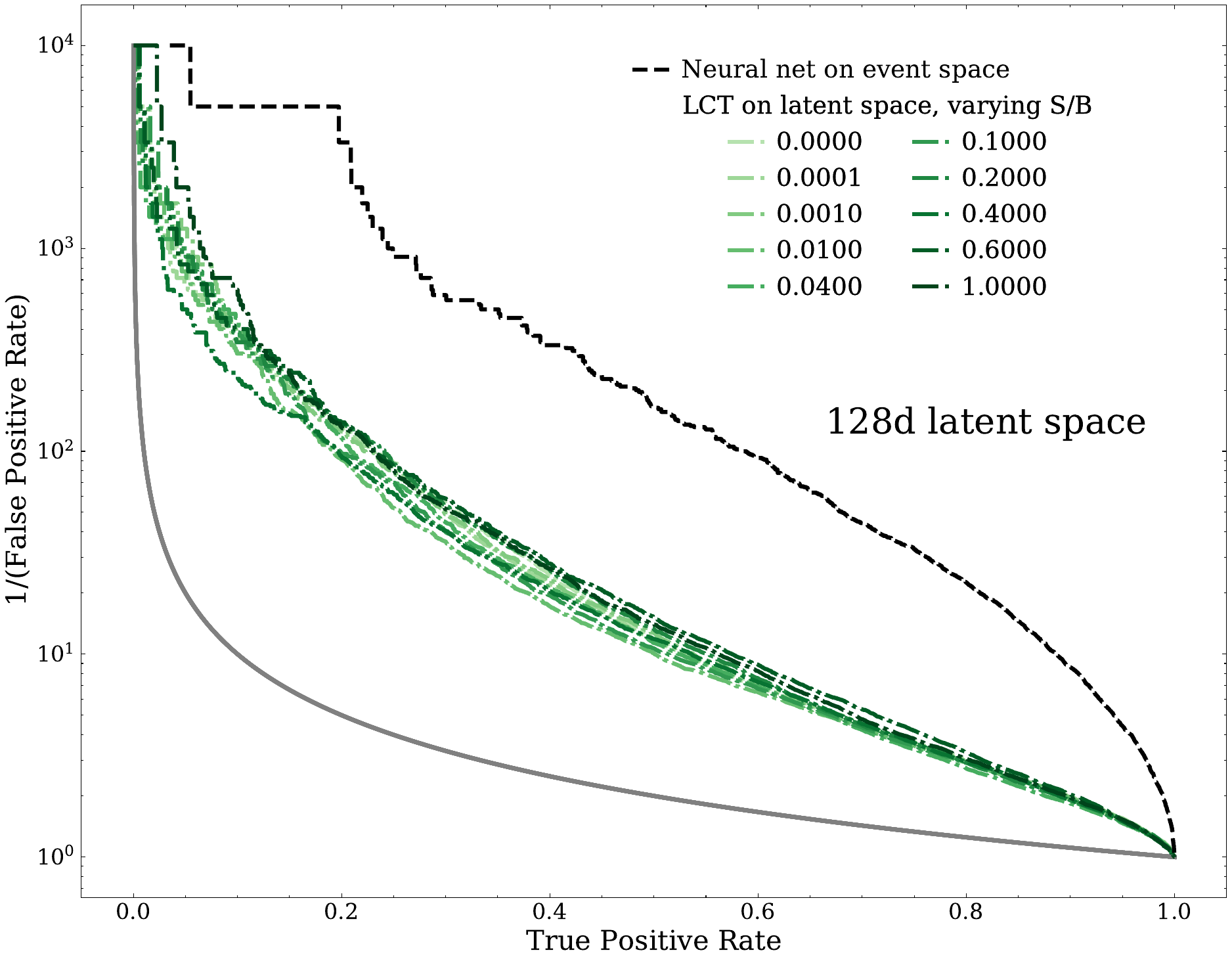}
    \caption{128d latent space}
    \label{fig:SvB_128_LCT}
  \end{subfigure}
  \begin{subfigure}{.49\linewidth}
    \includegraphics[width=\SingleImageWidth\columnwidth]{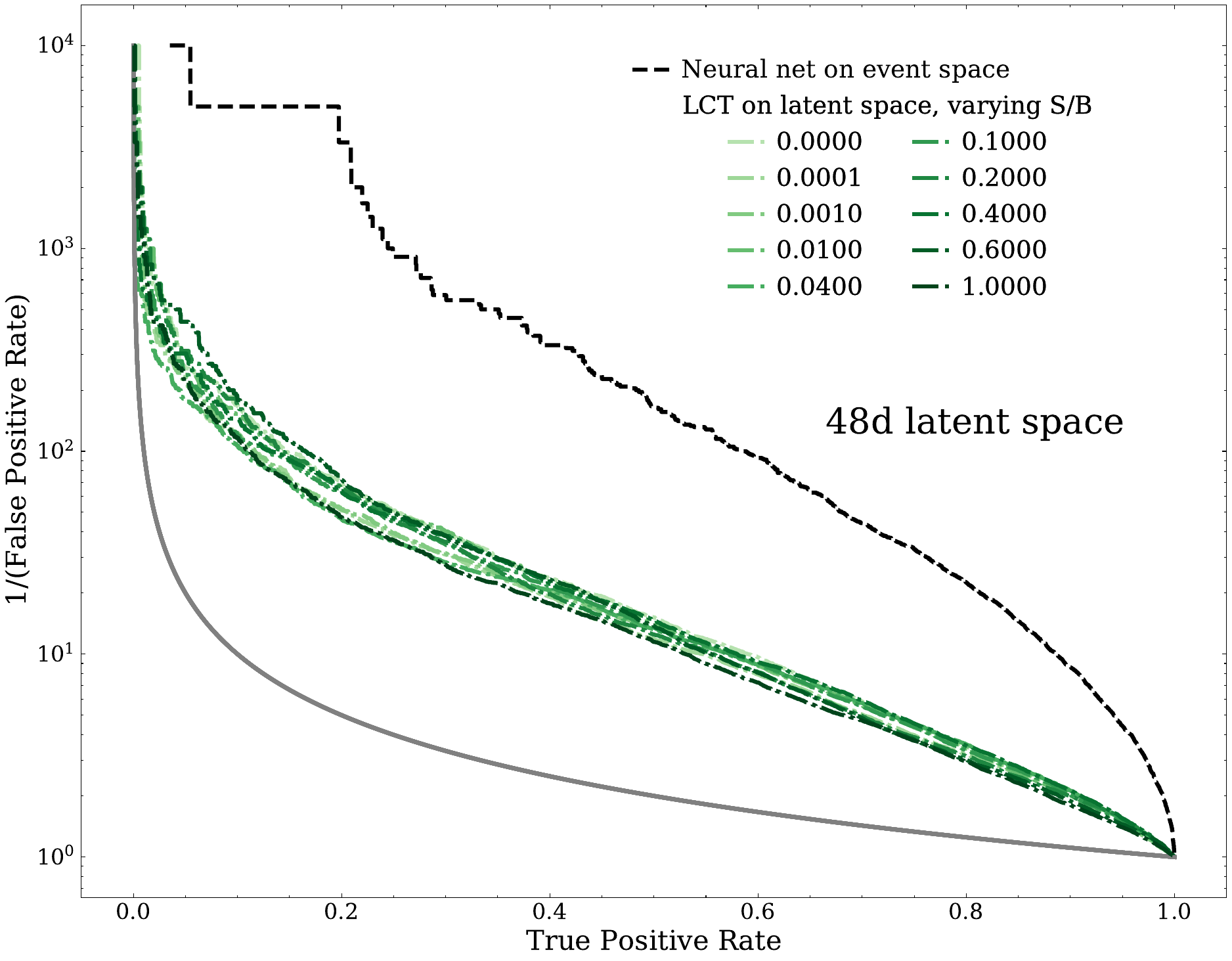}
    \caption{48d latent space}
    \label{fig:SvB_48_LCT}
  \end{subfigure}
  \caption{Classifier efficiency curves for a latent space linear classifier test trained on varying amounts of signal fraction. The
classifier performance is again robust with respect to the signal fraction of the training data.}
  \label{fig:SvB_LCT}
\end{figure*}

In this section, we provide the analogues to \Fig{fig:fpr_tpr_dim_scan} and \Fig{fig:SvB} with the transformer-encoder efficiency curves calculated for a binary linear classifier test run on the latent space representations (rather than a binary FCN). This aligns with the field-standard way to evaluate representations of jets, through a LCT. However, these plots (\Fig{fig:fpr_tpr_dim_scan_LCT} and \Fig{fig:SvB_LCT}) are not shown in the main text of this report as a realistic anomaly detection analysis would be carried out using fully connected networks.

\section{How data-hungry are the neural networks?}
\label{sec:datahungry}

In this section, we provide plots illustrating the data-hungry nature of the transformer-encoder network. The performance of the binary classifier trained on the latent space dijet representations, as shown in \Fig{fig:cwola}, is admittedly low. However, it is likely that the performance could improve if the classifiers were trained on a larger amount of data.

\begin{figure*}
  \centering
 
  \begin{subfigure}{.49\linewidth}
    \includegraphics[width=\SingleImageWidth\columnwidth]{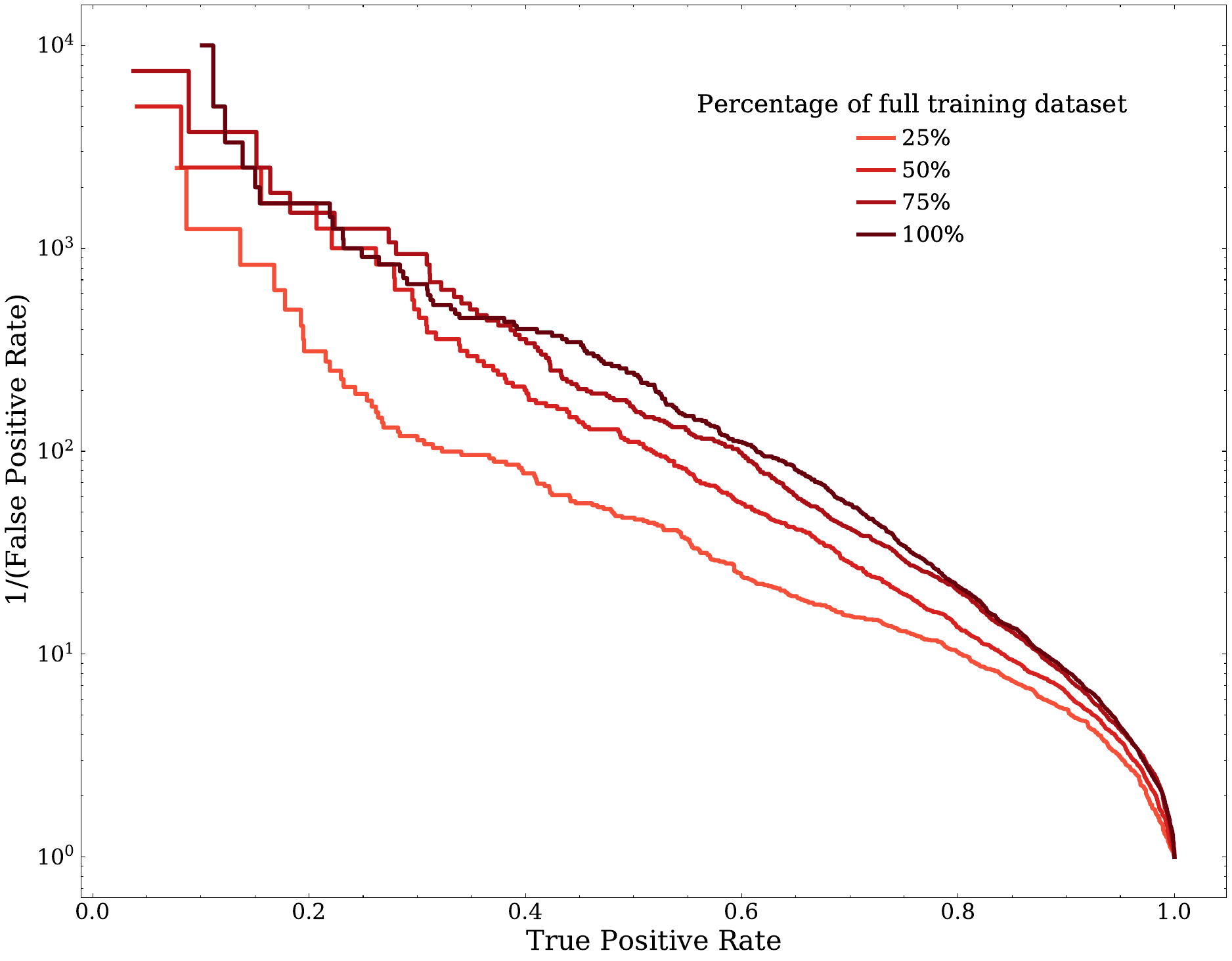}
    \caption{FCN network architecture. The network performance shows only a mild degree of saturation.}
    \label{fig:FCN_data_fraction}
  \end{subfigure}
   \begin{subfigure}{.49\linewidth}
    \includegraphics[width=\SingleImageWidth\columnwidth]{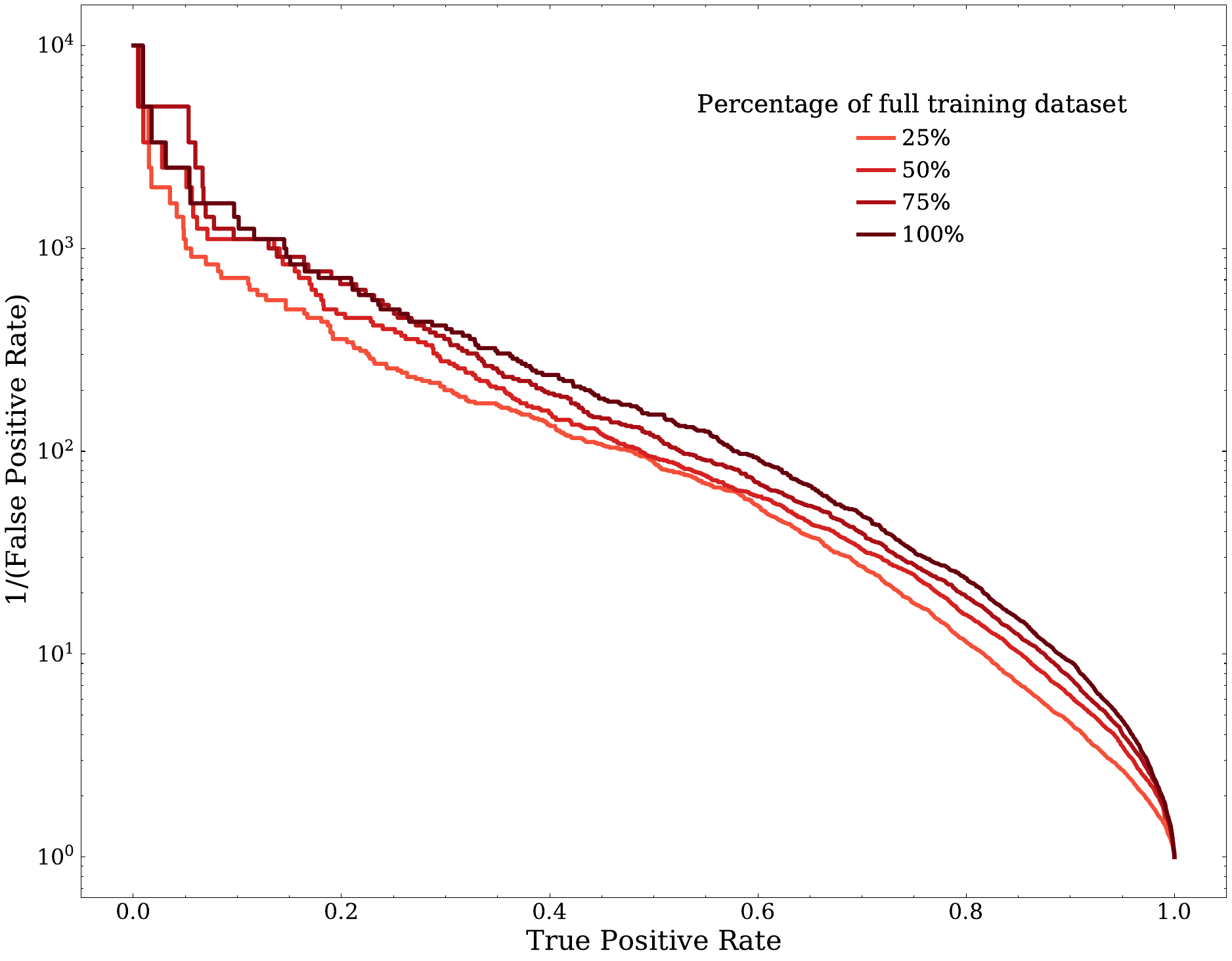}
    \caption{Trans+BC network architecture. The network performance steadily increases with amount of training data, showing no signs of saturation.}
    \label{fig:trans_BC_data_fraction}
  \end{subfigure}
  \caption{Classifier efficiency curves for a binary classifier trained on an increasing percentage of the full dijet dataset.}
  \label{fig:data_hungry}
\end{figure*}

In \Fig{fig:FCN_data_fraction}, we train a FCN on a varying fraction of the available dijet dataset (a 100\% training fraction makes use of all 85,000 signal and 85,000 background events). In \Fig{fig:trans_BC_data_fraction}, we repeat this procedure for a Trans+BC network. In both cases, the ROC AUCs of the trained binary classifiers do not appear to be saturated when trained on the full dataset.


\clearpage

\bibliography{main,HEPML}

\providecommand{\noopsort}[1]{}\providecommand{\singleletter}[1]{#1}%
\begin{thebibliography}{36}%
\makeatletter
\providecommand \@ifxundefined [1]{%
 \@ifx{#1\undefined}
}%
\providecommand \@ifnum [1]{%
 \ifnum #1\expandafter \@firstoftwo
 \else \expandafter \@secondoftwo
 \fi
}%
\providecommand \@ifx [1]{%
 \ifx #1\expandafter \@firstoftwo
 \else \expandafter \@secondoftwo
 \fi
}%
\providecommand \natexlab [1]{#1}%
\providecommand \enquote  [1]{``#1''}%
\providecommand \bibnamefont  [1]{#1}%
\providecommand \bibfnamefont [1]{#1}%
\providecommand \citenamefont [1]{#1}%
\providecommand \href@noop [0]{\@secondoftwo}%
\providecommand \href [0]{\begingroup \@sanitize@url \@href}%
\providecommand \@href[1]{\@@startlink{#1}\@@href}%
\providecommand \@@href[1]{\endgroup#1\@@endlink}%
\providecommand \@sanitize@url [0]{\catcode `\\12\catcode `\$12\catcode
  `\&12\catcode `\#12\catcode `\^12\catcode `\_12\catcode `\%12\relax}%
\providecommand \@@startlink[1]{}%
\providecommand \@@endlink[0]{}%
\providecommand \url  [0]{\begingroup\@sanitize@url \@url }%
\providecommand \@url [1]{\endgroup\@href {#1}{\urlprefix }}%
\providecommand \urlprefix  [0]{URL }%
\providecommand \Eprint [0]{\href }%
\providecommand \doibase [0]{http://dx.doi.org/}%
\providecommand \selectlanguage [0]{\@gobble}%
\providecommand \bibinfo  [0]{\@secondoftwo}%
\providecommand \bibfield  [0]{\@secondoftwo}%
\providecommand \translation [1]{[#1]}%
\providecommand \BibitemOpen [0]{}%
\providecommand \bibitemStop [0]{}%
\providecommand \bibitemNoStop [0]{.\EOS\space}%
\providecommand \EOS [0]{\spacefactor3000\relax}%
\providecommand \BibitemShut  [1]{\csname bibitem#1\endcsname}%
\let\auto@bib@innerbib\@empty
\bibitem [{\citenamefont {Craig}\ \emph {et~al.}(2019)\citenamefont {Craig},
  \citenamefont {Draper}, \citenamefont {Kong}, \citenamefont {Ng},\ and\
  \citenamefont {Whiteson}}]{Craig:2016rqv}%
  \BibitemOpen
  \bibfield  {author} {\bibinfo {author} {\bibfnamefont {Nathaniel}\
  \bibnamefont {Craig}}, \bibinfo {author} {\bibfnamefont {Patrick}\
  \bibnamefont {Draper}}, \bibinfo {author} {\bibfnamefont {Kyoungchul}\
  \bibnamefont {Kong}}, \bibinfo {author} {\bibfnamefont {Yvonne}\ \bibnamefont
  {Ng}}, \ and\ \bibinfo {author} {\bibfnamefont {Daniel}\ \bibnamefont
  {Whiteson}},\ }\bibfield  {title} {\enquote {\bibinfo {title} {{The
  unexplored landscape of two-body resonances}},}\ }\href {\doibase
  10.5506/APhysPolB.50.837} {\bibfield  {journal} {\bibinfo  {journal} {Acta
  Phys. Polon. B}\ }\textbf {\bibinfo {volume} {50}},\ \bibinfo {pages} {837}
  (\bibinfo {year} {2019})},\ \Eprint {http://arxiv.org/abs/1610.09392}
  {arXiv:1610.09392 [hep-ph]} \BibitemShut {NoStop}%
\bibitem [{\citenamefont {Kim}\ \emph {et~al.}(2020)\citenamefont {Kim},
  \citenamefont {Kong}, \citenamefont {Nachman},\ and\ \citenamefont
  {Whiteson}}]{Kim:2019rhy}%
  \BibitemOpen
  \bibfield  {author} {\bibinfo {author} {\bibfnamefont {Jeong~Han}\
  \bibnamefont {Kim}}, \bibinfo {author} {\bibfnamefont {Kyoungchul}\
  \bibnamefont {Kong}}, \bibinfo {author} {\bibfnamefont {Benjamin}\
  \bibnamefont {Nachman}}, \ and\ \bibinfo {author} {\bibfnamefont {Daniel}\
  \bibnamefont {Whiteson}},\ }\bibfield  {title} {\enquote {\bibinfo {title}
  {{The motivation and status of two-body resonance decays after the LHC Run 2
  and beyond}},}\ }\href {\doibase 10.1007/JHEP04(2020)030} {\bibfield
  {journal} {\bibinfo  {journal} {JHEP}\ }\textbf {\bibinfo {volume} {04}},\
  \bibinfo {pages} {030} (\bibinfo {year} {2020})},\ \Eprint
  {http://arxiv.org/abs/1907.06659} {arXiv:1907.06659 [hep-ph]} \BibitemShut
  {NoStop}%
\bibitem [{\citenamefont {Kasieczka}\ \emph {et~al.}(2021)\citenamefont
  {Kasieczka} \emph {et~al.}}]{Kasieczka:2021xcg}%
  \BibitemOpen
  \bibfield  {author} {\bibinfo {author} {\bibfnamefont {Gregor}\ \bibnamefont
  {Kasieczka}} \emph {et~al.},\ }\bibfield  {title} {\enquote {\bibinfo {title}
  {{The LHC Olympics 2020 a community challenge for anomaly detection in high
  energy physics}},}\ }\href {\doibase 10.1088/1361-6633/ac36b9} {\bibfield
  {journal} {\bibinfo  {journal} {Rept. Prog. Phys.}\ }\textbf {\bibinfo
  {volume} {84}},\ \bibinfo {pages} {124201} (\bibinfo {year} {2021})},\
  \Eprint {http://arxiv.org/abs/2101.08320} {arXiv:2101.08320 [hep-ph]}
  \BibitemShut {NoStop}%
\bibitem [{\citenamefont {Aarrestad}\ \emph {et~al.}(2021)\citenamefont
  {Aarrestad} \emph {et~al.}}]{Aarrestad:2021oeb}%
  \BibitemOpen
  \bibfield  {author} {\bibinfo {author} {\bibfnamefont {T.}~\bibnamefont
  {Aarrestad}} \emph {et~al.},\ }\bibfield  {title} {\enquote {\bibinfo {title}
  {{The Dark Machines Anomaly Score Challenge: Benchmark Data and Model
  Independent Event Classification for the Large Hadron Collider}},}\
  }\href@noop {} {\  (\bibinfo {year} {2021})},\ \Eprint
  {http://arxiv.org/abs/2105.14027} {arXiv:2105.14027 [hep-ph]} \BibitemShut
  {NoStop}%
\bibitem [{\citenamefont {Karagiorgi}\ \emph {et~al.}(2021)\citenamefont
  {Karagiorgi}, \citenamefont {Kasieczka}, \citenamefont {Kravitz},
  \citenamefont {Nachman},\ and\ \citenamefont {Shih}}]{Karagiorgi:2021ngt}%
  \BibitemOpen
  \bibfield  {author} {\bibinfo {author} {\bibfnamefont {Georgia}\ \bibnamefont
  {Karagiorgi}}, \bibinfo {author} {\bibfnamefont {Gregor}\ \bibnamefont
  {Kasieczka}}, \bibinfo {author} {\bibfnamefont {Scott}\ \bibnamefont
  {Kravitz}}, \bibinfo {author} {\bibfnamefont {Benjamin}\ \bibnamefont
  {Nachman}}, \ and\ \bibinfo {author} {\bibfnamefont {David}\ \bibnamefont
  {Shih}},\ }\bibfield  {title} {\enquote {\bibinfo {title} {{Machine Learning
  in the Search for New Fundamental Physics}},}\ }\href@noop {} {\  (\bibinfo
  {year} {2021})},\ \Eprint {http://arxiv.org/abs/2112.03769} {arXiv:2112.03769
  [hep-ph]} \BibitemShut {NoStop}%
\bibitem [{\citenamefont {Feickert}\ and\ \citenamefont
  {Nachman}(2021)}]{Feickert:2021ajf}%
  \BibitemOpen
  \bibfield  {author} {\bibinfo {author} {\bibfnamefont {Matthew}\ \bibnamefont
  {Feickert}}\ and\ \bibinfo {author} {\bibfnamefont {Benjamin}\ \bibnamefont
  {Nachman}},\ }\bibfield  {title} {\enquote {\bibinfo {title} {{A Living
  Review of Machine Learning for Particle Physics}},}\ }\href@noop {} {\
  (\bibinfo {year} {2021})},\ \Eprint {http://arxiv.org/abs/2102.02770}
  {arXiv:2102.02770 [hep-ph]} \BibitemShut {NoStop}%
\bibitem [{\citenamefont {{ATLAS
  Collaboration}}(2020)}]{collaboration2020dijet}%
  \BibitemOpen
  \bibfield  {author} {\bibinfo {author} {\bibnamefont {{ATLAS
  Collaboration}}},\ }\bibfield  {title} {\enquote {\bibinfo {title} {{Dijet
  resonance search with weak supervision using 13 TeV pp collisions in the
  ATLAS detector}},}\ }\href {\doibase 10.1103/PhysRevLett.125.131801} {\
  (\bibinfo {year} {2020}),\ 10.1103/PhysRevLett.125.131801},\ \Eprint
  {http://arxiv.org/abs/2005.02983} {arXiv:2005.02983 [hep-ex]} \BibitemShut
  {NoStop}%
\bibitem [{\citenamefont {Hajer}\ \emph {et~al.}(2018)\citenamefont {Hajer},
  \citenamefont {Li}, \citenamefont {Liu},\ and\ \citenamefont
  {Wang}}]{Hajer:2018kqm}%
  \BibitemOpen
  \bibfield  {author} {\bibinfo {author} {\bibfnamefont {Jan}\ \bibnamefont
  {Hajer}}, \bibinfo {author} {\bibfnamefont {Ying-Ying}\ \bibnamefont {Li}},
  \bibinfo {author} {\bibfnamefont {Tao}\ \bibnamefont {Liu}}, \ and\ \bibinfo
  {author} {\bibfnamefont {He}~\bibnamefont {Wang}},\ }\bibfield  {title}
  {\enquote {\bibinfo {title} {{Novelty Detection Meets Collider Physics}},}\
  }\href {\doibase 10.1103/PhysRevD.101.076015} {\  (\bibinfo {year} {2018}),\
  10.1103/PhysRevD.101.076015},\ \Eprint {http://arxiv.org/abs/1807.10261}
  {arXiv:1807.10261 [hep-ph]} \BibitemShut {NoStop}%
\bibitem [{\citenamefont {Farina}\ \emph {et~al.}(2018)\citenamefont {Farina},
  \citenamefont {Nakai},\ and\ \citenamefont {Shih}}]{Farina:2018fyg}%
  \BibitemOpen
  \bibfield  {author} {\bibinfo {author} {\bibfnamefont {Marco}\ \bibnamefont
  {Farina}}, \bibinfo {author} {\bibfnamefont {Yuichiro}\ \bibnamefont
  {Nakai}}, \ and\ \bibinfo {author} {\bibfnamefont {David}\ \bibnamefont
  {Shih}},\ }\bibfield  {title} {\enquote {\bibinfo {title} {{Searching for New
  Physics with Deep Autoencoders}},}\ }\href {\doibase
  10.1103/PhysRevD.101.075021} {\  (\bibinfo {year} {2018}),\
  10.1103/PhysRevD.101.075021},\ \Eprint {http://arxiv.org/abs/1808.08992}
  {arXiv:1808.08992 [hep-ph]} \BibitemShut {NoStop}%
\bibitem [{\citenamefont {Heimel}\ \emph {et~al.}(2019)\citenamefont {Heimel},
  \citenamefont {Kasieczka}, \citenamefont {Plehn},\ and\ \citenamefont
  {Thompson}}]{Heimel:2018mkt}%
  \BibitemOpen
  \bibfield  {author} {\bibinfo {author} {\bibfnamefont {Theo}\ \bibnamefont
  {Heimel}}, \bibinfo {author} {\bibfnamefont {Gregor}\ \bibnamefont
  {Kasieczka}}, \bibinfo {author} {\bibfnamefont {Tilman}\ \bibnamefont
  {Plehn}}, \ and\ \bibinfo {author} {\bibfnamefont {Jennifer~M.}\ \bibnamefont
  {Thompson}},\ }\bibfield  {title} {\enquote {\bibinfo {title} {{QCD or
  What?}}}\ }\href {\doibase 10.21468/SciPostPhys.6.3.030} {\bibfield
  {journal} {\bibinfo  {journal} {SciPost Phys.}\ }\textbf {\bibinfo {volume}
  {6}},\ \bibinfo {pages} {030} (\bibinfo {year} {2019})},\ \Eprint
  {http://arxiv.org/abs/1808.08979} {arXiv:1808.08979 [hep-ph]} \BibitemShut
  {NoStop}%
\bibitem [{\citenamefont {Bortolato}\ \emph {et~al.}(2021)\citenamefont
  {Bortolato}, \citenamefont {Dillon}, \citenamefont {Kamenik},\ and\
  \citenamefont {Smolkovi\v{c}}}]{Bortolato:2021zic}%
  \BibitemOpen
  \bibfield  {author} {\bibinfo {author} {\bibfnamefont {Bla\v{z}}\
  \bibnamefont {Bortolato}}, \bibinfo {author} {\bibfnamefont {Barry~M.}\
  \bibnamefont {Dillon}}, \bibinfo {author} {\bibfnamefont {Jernej~F.}\
  \bibnamefont {Kamenik}}, \ and\ \bibinfo {author} {\bibfnamefont {Aleks}\
  \bibnamefont {Smolkovi\v{c}}},\ }\bibfield  {title} {\enquote {\bibinfo
  {title} {{Bump Hunting in Latent Space}},}\ }\href@noop {} {\  (\bibinfo
  {year} {2021})},\ \Eprint {http://arxiv.org/abs/2103.06595} {arXiv:2103.06595
  [hep-ph]} \BibitemShut {NoStop}%
\bibitem [{\citenamefont {Hayat}\ \emph {et~al.}(2020)\citenamefont {Hayat},
  \citenamefont {Stein}, \citenamefont {Harrington}, \citenamefont {Lukić},\
  and\ \citenamefont {Mustafa}}]{https://doi.org/10.48550/arxiv.2012.13083}%
  \BibitemOpen
  \bibfield  {author} {\bibinfo {author} {\bibfnamefont {Md~Abul}\ \bibnamefont
  {Hayat}}, \bibinfo {author} {\bibfnamefont {George}\ \bibnamefont {Stein}},
  \bibinfo {author} {\bibfnamefont {Peter}\ \bibnamefont {Harrington}},
  \bibinfo {author} {\bibfnamefont {Zarija}\ \bibnamefont {Lukić}}, \ and\
  \bibinfo {author} {\bibfnamefont {Mustafa}\ \bibnamefont {Mustafa}},\ }\href
  {\doibase 10.48550/ARXIV.2012.13083} {\enquote {\bibinfo {title}
  {Self-supervised representation learning for astronomical images},}\ }
  (\bibinfo {year} {2020})\BibitemShut {NoStop}%
\bibitem [{\citenamefont {Dillon}\ \emph {et~al.}(2021)\citenamefont {Dillon},
  \citenamefont {Kasieczka}, \citenamefont {Olischlager}, \citenamefont
  {Plehn}, \citenamefont {Sorrenson},\ and\ \citenamefont
  {Vogel}}]{Dillon:2021gag}%
  \BibitemOpen
  \bibfield  {author} {\bibinfo {author} {\bibfnamefont {Barry~M.}\
  \bibnamefont {Dillon}}, \bibinfo {author} {\bibfnamefont {Gregor}\
  \bibnamefont {Kasieczka}}, \bibinfo {author} {\bibfnamefont {Hans}\
  \bibnamefont {Olischlager}}, \bibinfo {author} {\bibfnamefont {Tilman}\
  \bibnamefont {Plehn}}, \bibinfo {author} {\bibfnamefont {Peter}\ \bibnamefont
  {Sorrenson}}, \ and\ \bibinfo {author} {\bibfnamefont {Lorenz}\ \bibnamefont
  {Vogel}},\ }\bibfield  {title} {\enquote {\bibinfo {title} {{Symmetries,
  Safety, and Self-Supervision}},}\ }\href@noop {} {\  (\bibinfo {year}
  {2021})},\ \Eprint {http://arxiv.org/abs/2108.04253} {arXiv:2108.04253
  [hep-ph]} \BibitemShut {NoStop}%
\bibitem [{\citenamefont {He}\ \emph {et~al.}(2015)\citenamefont {He},
  \citenamefont {Zhang}, \citenamefont {Ren},\ and\ \citenamefont
  {Sun}}]{https://doi.org/10.48550/arxiv.1512.03385}%
  \BibitemOpen
  \bibfield  {author} {\bibinfo {author} {\bibfnamefont {Kaiming}\ \bibnamefont
  {He}}, \bibinfo {author} {\bibfnamefont {Xiangyu}\ \bibnamefont {Zhang}},
  \bibinfo {author} {\bibfnamefont {Shaoqing}\ \bibnamefont {Ren}}, \ and\
  \bibinfo {author} {\bibfnamefont {Jian}\ \bibnamefont {Sun}},\ }\href
  {\doibase 10.48550/ARXIV.1512.03385} {\enquote {\bibinfo {title} {Deep
  residual learning for image recognition},}\ } (\bibinfo {year}
  {2015})\BibitemShut {NoStop}%
\bibitem [{\citenamefont {Metodiev}\ \emph {et~al.}(2017)\citenamefont
  {Metodiev}, \citenamefont {Nachman},\ and\ \citenamefont
  {Thaler}}]{Metodiev:2017vrx}%
  \BibitemOpen
  \bibfield  {author} {\bibinfo {author} {\bibfnamefont {Eric~M.}\ \bibnamefont
  {Metodiev}}, \bibinfo {author} {\bibfnamefont {Benjamin}\ \bibnamefont
  {Nachman}}, \ and\ \bibinfo {author} {\bibfnamefont {Jesse}\ \bibnamefont
  {Thaler}},\ }\bibfield  {title} {\enquote {\bibinfo {title} {{Classification
  without labels: Learning from mixed samples in high energy physics}},}\
  }\href {\doibase 10.1007/JHEP10(2017)174} {\bibfield  {journal} {\bibinfo
  {journal} {JHEP}\ }\textbf {\bibinfo {volume} {10}},\ \bibinfo {pages} {174}
  (\bibinfo {year} {2017})},\ \Eprint {http://arxiv.org/abs/1708.02949}
  {arXiv:1708.02949 [hep-ph]} \BibitemShut {NoStop}%
\bibitem [{\citenamefont {Collins}\ \emph {et~al.}(2018)\citenamefont
  {Collins}, \citenamefont {Howe},\ and\ \citenamefont
  {Nachman}}]{Collins:2018epr}%
  \BibitemOpen
  \bibfield  {author} {\bibinfo {author} {\bibfnamefont {Jack~H.}\ \bibnamefont
  {Collins}}, \bibinfo {author} {\bibfnamefont {Kiel}\ \bibnamefont {Howe}}, \
  and\ \bibinfo {author} {\bibfnamefont {Benjamin}\ \bibnamefont {Nachman}},\
  }\bibfield  {title} {\enquote {\bibinfo {title} {{Anomaly Detection for
  Resonant New Physics with Machine Learning}},}\ }\href {\doibase
  10.1103/PhysRevLett.121.241803} {\bibfield  {journal} {\bibinfo  {journal}
  {Phys. Rev. Lett.}\ }\textbf {\bibinfo {volume} {121}},\ \bibinfo {pages}
  {241803} (\bibinfo {year} {2018})},\ \Eprint
  {http://arxiv.org/abs/1805.02664} {arXiv:1805.02664 [hep-ph]} \BibitemShut
  {NoStop}%
\bibitem [{\citenamefont {Collins}\ \emph {et~al.}(2019)\citenamefont
  {Collins}, \citenamefont {Howe},\ and\ \citenamefont
  {Nachman}}]{Collins:2019jip}%
  \BibitemOpen
  \bibfield  {author} {\bibinfo {author} {\bibfnamefont {Jack~H.}\ \bibnamefont
  {Collins}}, \bibinfo {author} {\bibfnamefont {Kiel}\ \bibnamefont {Howe}}, \
  and\ \bibinfo {author} {\bibfnamefont {Benjamin}\ \bibnamefont {Nachman}},\
  }\bibfield  {title} {\enquote {\bibinfo {title} {{Extending the search for
  new resonances with machine learning}},}\ }\href {\doibase
  10.1103/PhysRevD.99.014038} {\bibfield  {journal} {\bibinfo  {journal} {Phys.
  Rev.}\ }\textbf {\bibinfo {volume} {D99}},\ \bibinfo {pages} {014038}
  (\bibinfo {year} {2019})},\ \Eprint {http://arxiv.org/abs/1902.02634}
  {arXiv:1902.02634 [hep-ph]} \BibitemShut {NoStop}%
\bibitem [{\citenamefont {Kasieczka}\ \emph {et~al.}(2019)\citenamefont
  {Kasieczka}, \citenamefont {Nachman},\ and\ \citenamefont
  {Shih}}]{LHCOlympics}%
  \BibitemOpen
  \bibfield  {author} {\bibinfo {author} {\bibfnamefont {Gregor}\ \bibnamefont
  {Kasieczka}}, \bibinfo {author} {\bibfnamefont {Benjamin}\ \bibnamefont
  {Nachman}}, \ and\ \bibinfo {author} {\bibfnamefont {David}\ \bibnamefont
  {Shih}},\ }\href@noop {} {\enquote {\bibinfo {title} {{Official Datasets for
  LHC Olympics 2020 Anomaly Detection Challenge (Version v6) [Data set].}}}\ }
  (\bibinfo {year} {2019}),\ \bibinfo {note}
  {https://doi.org/10.5281/zenodo.4536624}\BibitemShut {NoStop}%
\bibitem [{\citenamefont {Chen}\ \emph {et~al.}(2020)\citenamefont {Chen},
  \citenamefont {Kornblith}, \citenamefont {Norouzi},\ and\ \citenamefont
  {Hinton}}]{chen2020simple}%
  \BibitemOpen
  \bibfield  {author} {\bibinfo {author} {\bibfnamefont {Ting}\ \bibnamefont
  {Chen}}, \bibinfo {author} {\bibfnamefont {Simon}\ \bibnamefont {Kornblith}},
  \bibinfo {author} {\bibfnamefont {Mohammad}\ \bibnamefont {Norouzi}}, \ and\
  \bibinfo {author} {\bibfnamefont {Geoffrey}\ \bibnamefont {Hinton}},\
  }\href@noop {} {\enquote {\bibinfo {title} {A simple framework for
  contrastive learning of visual representations},}\ } (\bibinfo {year}
  {2020}),\ \Eprint {http://arxiv.org/abs/2002.05709} {arXiv:2002.05709
  [cs.LG]} \BibitemShut {NoStop}%
\bibitem [{\citenamefont {Sjostrand}\ \emph {et~al.}(2006)\citenamefont
  {Sjostrand}, \citenamefont {Mrenna},\ and\ \citenamefont
  {Skands}}]{Sjostrand:2006za}%
  \BibitemOpen
  \bibfield  {author} {\bibinfo {author} {\bibfnamefont {Torbjorn}\
  \bibnamefont {Sjostrand}}, \bibinfo {author} {\bibfnamefont {Stephen}\
  \bibnamefont {Mrenna}}, \ and\ \bibinfo {author} {\bibfnamefont {Peter~Z.}\
  \bibnamefont {Skands}},\ }\bibfield  {title} {\enquote {\bibinfo {title}
  {{PYTHIA 6.4 Physics and Manual}},}\ }\href {\doibase
  10.1088/1126-6708/2006/05/026} {\bibfield  {journal} {\bibinfo  {journal}
  {JHEP}\ }\textbf {\bibinfo {volume} {05}},\ \bibinfo {pages} {026} (\bibinfo
  {year} {2006})},\ \Eprint {http://arxiv.org/abs/hep-ph/0603175}
  {arXiv:hep-ph/0603175} \BibitemShut {NoStop}%
\bibitem [{\citenamefont {Sj\"ostrand}\ \emph {et~al.}(2015)\citenamefont
  {Sj\"ostrand}, \citenamefont {Ask}, \citenamefont {Christiansen},
  \citenamefont {Corke}, \citenamefont {Desai}, \citenamefont {Ilten},
  \citenamefont {Mrenna}, \citenamefont {Prestel}, \citenamefont {Rasmussen},\
  and\ \citenamefont {Skands}}]{Sjostrand:2014zea}%
  \BibitemOpen
  \bibfield  {author} {\bibinfo {author} {\bibfnamefont {Torbj\"orn}\
  \bibnamefont {Sj\"ostrand}}, \bibinfo {author} {\bibfnamefont {Stefan}\
  \bibnamefont {Ask}}, \bibinfo {author} {\bibfnamefont {Jesper~R.}\
  \bibnamefont {Christiansen}}, \bibinfo {author} {\bibfnamefont {Richard}\
  \bibnamefont {Corke}}, \bibinfo {author} {\bibfnamefont {Nishita}\
  \bibnamefont {Desai}}, \bibinfo {author} {\bibfnamefont {Philip}\
  \bibnamefont {Ilten}}, \bibinfo {author} {\bibfnamefont {Stephen}\
  \bibnamefont {Mrenna}}, \bibinfo {author} {\bibfnamefont {Stefan}\
  \bibnamefont {Prestel}}, \bibinfo {author} {\bibfnamefont {Christine~O.}\
  \bibnamefont {Rasmussen}}, \ and\ \bibinfo {author} {\bibfnamefont
  {Peter~Z.}\ \bibnamefont {Skands}},\ }\bibfield  {title} {\enquote {\bibinfo
  {title} {{An introduction to PYTHIA 8.2}},}\ }\href {\doibase
  10.1016/j.cpc.2015.01.024} {\bibfield  {journal} {\bibinfo  {journal}
  {Comput. Phys. Commun.}\ }\textbf {\bibinfo {volume} {191}},\ \bibinfo
  {pages} {159--177} (\bibinfo {year} {2015})},\ \Eprint
  {http://arxiv.org/abs/1410.3012} {arXiv:1410.3012 [hep-ph]} \BibitemShut
  {NoStop}%
\bibitem [{\citenamefont {de~Favereau}\ \emph {et~al.}(2014)\citenamefont
  {de~Favereau}, \citenamefont {Delaere}, \citenamefont {Demin}, \citenamefont
  {Giammanco}, \citenamefont {Lema\^\i{}tre}, \citenamefont {Mertens},\ and\
  \citenamefont {Selvaggi}}]{deFavereau:2013fsa}%
  \BibitemOpen
  \bibfield  {author} {\bibinfo {author} {\bibfnamefont {J.}~\bibnamefont
  {de~Favereau}}, \bibinfo {author} {\bibfnamefont {C.}~\bibnamefont
  {Delaere}}, \bibinfo {author} {\bibfnamefont {P.}~\bibnamefont {Demin}},
  \bibinfo {author} {\bibfnamefont {A.}~\bibnamefont {Giammanco}}, \bibinfo
  {author} {\bibfnamefont {V.}~\bibnamefont {Lema\^\i{}tre}}, \bibinfo {author}
  {\bibfnamefont {A.}~\bibnamefont {Mertens}}, \ and\ \bibinfo {author}
  {\bibfnamefont {M.}~\bibnamefont {Selvaggi}} (\bibinfo {collaboration}
  {DELPHES 3}),\ }\bibfield  {title} {\enquote {\bibinfo {title} {{DELPHES 3, A
  modular framework for fast simulation of a generic collider experiment}},}\
  }\href {\doibase 10.1007/JHEP02(2014)057} {\bibfield  {journal} {\bibinfo
  {journal} {JHEP}\ }\textbf {\bibinfo {volume} {02}},\ \bibinfo {pages} {057}
  (\bibinfo {year} {2014})},\ \Eprint {http://arxiv.org/abs/1307.6346}
  {arXiv:1307.6346 [hep-ex]} \BibitemShut {NoStop}%
\bibitem [{\citenamefont {Cacciari}\ \emph {et~al.}(2012)\citenamefont
  {Cacciari}, \citenamefont {Salam},\ and\ \citenamefont
  {Soyez}}]{Cacciari:2011ma}%
  \BibitemOpen
  \bibfield  {author} {\bibinfo {author} {\bibfnamefont {Matteo}\ \bibnamefont
  {Cacciari}}, \bibinfo {author} {\bibfnamefont {Gavin~P.}\ \bibnamefont
  {Salam}}, \ and\ \bibinfo {author} {\bibfnamefont {Gregory}\ \bibnamefont
  {Soyez}},\ }\bibfield  {title} {\enquote {\bibinfo {title} {{FastJet User
  Manual}},}\ }\href {\doibase 10.1140/epjc/s10052-012-1896-2} {\bibfield
  {journal} {\bibinfo  {journal} {Eur. Phys. J. C}\ }\textbf {\bibinfo {volume}
  {72}},\ \bibinfo {pages} {1896} (\bibinfo {year} {2012})},\ \Eprint
  {http://arxiv.org/abs/1111.6097} {arXiv:1111.6097 [hep-ph]} \BibitemShut
  {NoStop}%
\bibitem [{\citenamefont {Cacciari}\ and\ \citenamefont
  {Salam}(2006)}]{Cacciari:2005hq}%
  \BibitemOpen
  \bibfield  {author} {\bibinfo {author} {\bibfnamefont {Matteo}\ \bibnamefont
  {Cacciari}}\ and\ \bibinfo {author} {\bibfnamefont {Gavin~P.}\ \bibnamefont
  {Salam}},\ }\bibfield  {title} {\enquote {\bibinfo {title} {{Dispelling the
  $N^{3}$ myth for the $k_t$ jet-finder}},}\ }\href {\doibase
  10.1016/j.physletb.2006.08.037} {\bibfield  {journal} {\bibinfo  {journal}
  {Phys. Lett. B}\ }\textbf {\bibinfo {volume} {641}},\ \bibinfo {pages}
  {57--61} (\bibinfo {year} {2006})},\ \Eprint
  {http://arxiv.org/abs/hep-ph/0512210} {arXiv:hep-ph/0512210} \BibitemShut
  {NoStop}%
\bibitem [{\citenamefont {Paszke}\ \emph {et~al.}(2019)\citenamefont {Paszke},
  \citenamefont {Gross}, \citenamefont {Massa}, \citenamefont {Lerer},
  \citenamefont {Bradbury}, \citenamefont {Chanan}, \citenamefont {Killeen},
  \citenamefont {Lin}, \citenamefont {Gimelshein}, \citenamefont {Antiga},
  \citenamefont {Desmaison}, \citenamefont {Kopf}, \citenamefont {Yang},
  \citenamefont {DeVito}, \citenamefont {Raison}, \citenamefont {Tejani},
  \citenamefont {Chilamkurthy}, \citenamefont {Steiner}, \citenamefont {Fang},
  \citenamefont {Bai},\ and\ \citenamefont {Chintala}}]{NEURIPS2019_9015}%
  \BibitemOpen
  \bibfield  {author} {\bibinfo {author} {\bibfnamefont {Adam}\ \bibnamefont
  {Paszke}}, \bibinfo {author} {\bibfnamefont {Sam}\ \bibnamefont {Gross}},
  \bibinfo {author} {\bibfnamefont {Francisco}\ \bibnamefont {Massa}}, \bibinfo
  {author} {\bibfnamefont {Adam}\ \bibnamefont {Lerer}}, \bibinfo {author}
  {\bibfnamefont {James}\ \bibnamefont {Bradbury}}, \bibinfo {author}
  {\bibfnamefont {Gregory}\ \bibnamefont {Chanan}}, \bibinfo {author}
  {\bibfnamefont {Trevor}\ \bibnamefont {Killeen}}, \bibinfo {author}
  {\bibfnamefont {Zeming}\ \bibnamefont {Lin}}, \bibinfo {author}
  {\bibfnamefont {Natalia}\ \bibnamefont {Gimelshein}}, \bibinfo {author}
  {\bibfnamefont {Luca}\ \bibnamefont {Antiga}}, \bibinfo {author}
  {\bibfnamefont {Alban}\ \bibnamefont {Desmaison}}, \bibinfo {author}
  {\bibfnamefont {Andreas}\ \bibnamefont {Kopf}}, \bibinfo {author}
  {\bibfnamefont {Edward}\ \bibnamefont {Yang}}, \bibinfo {author}
  {\bibfnamefont {Zachary}\ \bibnamefont {DeVito}}, \bibinfo {author}
  {\bibfnamefont {Martin}\ \bibnamefont {Raison}}, \bibinfo {author}
  {\bibfnamefont {Alykhan}\ \bibnamefont {Tejani}}, \bibinfo {author}
  {\bibfnamefont {Sasank}\ \bibnamefont {Chilamkurthy}}, \bibinfo {author}
  {\bibfnamefont {Benoit}\ \bibnamefont {Steiner}}, \bibinfo {author}
  {\bibfnamefont {Lu}~\bibnamefont {Fang}}, \bibinfo {author} {\bibfnamefont
  {Junjie}\ \bibnamefont {Bai}}, \ and\ \bibinfo {author} {\bibfnamefont
  {Soumith}\ \bibnamefont {Chintala}},\ }\bibfield  {title} {\enquote {\bibinfo
  {title} {Pytorch: An imperative style, high-performance deep learning
  library},}\ }in\ \href
  {http://papers.neurips.cc/paper/9015-pytorch-an-imperative-style-high-performance-deep-learning-library.pdf}
  {\emph {\bibinfo {booktitle} {Advances in Neural Information Processing
  Systems 32}}}\ (\bibinfo  {publisher} {Curran Associates, Inc.},\ \bibinfo
  {year} {2019})\ pp.\ \bibinfo {pages} {8024--8035}\BibitemShut {NoStop}%
\bibitem [{\citenamefont {Kingma}\ and\ \citenamefont {Ba}(2014)}]{adam}%
  \BibitemOpen
  \bibfield  {author} {\bibinfo {author} {\bibfnamefont {Diederik~P.}\
  \bibnamefont {Kingma}}\ and\ \bibinfo {author} {\bibfnamefont {Jimmy}\
  \bibnamefont {Ba}},\ }\href {\doibase 10.48550/ARXIV.1412.6980} {\enquote
  {\bibinfo {title} {Adam: A method for stochastic optimization},}\ } (\bibinfo
  {year} {2014})\BibitemShut {NoStop}%
\bibitem [{\citenamefont {Pedregosa}\ \emph {et~al.}(2012)\citenamefont
  {Pedregosa} \emph {et~al.}}]{scikit}%
  \BibitemOpen
  \bibfield  {author} {\bibinfo {author} {\bibfnamefont {Fabian}\ \bibnamefont
  {Pedregosa}} \emph {et~al.},\ }\bibfield  {title} {\enquote {\bibinfo {title}
  {Scikit-learn: Machine learning in python},}\ }\href {\doibase
  10.48550/arXiv.1201.0490} {\  (\bibinfo {year} {2012}),\
  10.48550/arXiv.1201.0490}\BibitemShut {NoStop}%
\bibitem [{\citenamefont {D'Agnolo}\ and\ \citenamefont
  {Wulzer}(2019)}]{DAgnolo:2018cun}%
  \BibitemOpen
  \bibfield  {author} {\bibinfo {author} {\bibfnamefont {Raffaele~Tito}\
  \bibnamefont {D'Agnolo}}\ and\ \bibinfo {author} {\bibfnamefont {Andrea}\
  \bibnamefont {Wulzer}},\ }\bibfield  {title} {\enquote {\bibinfo {title}
  {{Learning New Physics from a Machine}},}\ }\href {\doibase
  10.1103/PhysRevD.99.015014} {\bibfield  {journal} {\bibinfo  {journal} {Phys.
  Rev.}\ }\textbf {\bibinfo {volume} {D99}},\ \bibinfo {pages} {015014}
  (\bibinfo {year} {2019})},\ \Eprint {http://arxiv.org/abs/1806.02350}
  {arXiv:1806.02350 [hep-ph]} \BibitemShut {NoStop}%
\bibitem [{\citenamefont {D'Agnolo}\ \emph {et~al.}(2019)\citenamefont
  {D'Agnolo}, \citenamefont {Grosso}, \citenamefont {Pierini}, \citenamefont
  {Wulzer},\ and\ \citenamefont {Zanetti}}]{DAgnolo:2019vbw}%
  \BibitemOpen
  \bibfield  {author} {\bibinfo {author} {\bibfnamefont {Raffaele~Tito}\
  \bibnamefont {D'Agnolo}}, \bibinfo {author} {\bibfnamefont {Gaia}\
  \bibnamefont {Grosso}}, \bibinfo {author} {\bibfnamefont {Maurizio}\
  \bibnamefont {Pierini}}, \bibinfo {author} {\bibfnamefont {Andrea}\
  \bibnamefont {Wulzer}}, \ and\ \bibinfo {author} {\bibfnamefont {Marco}\
  \bibnamefont {Zanetti}},\ }\bibfield  {title} {\enquote {\bibinfo {title}
  {{Learning Multivariate New Physics}},}\ }\href {\doibase
  10.1140/epjc/s10052-021-08853-y} {\  (\bibinfo {year} {2019}),\
  10.1140/epjc/s10052-021-08853-y},\ \Eprint {http://arxiv.org/abs/1912.12155}
  {arXiv:1912.12155 [hep-ph]} \BibitemShut {NoStop}%
\bibitem [{\citenamefont {d'Agnolo}\ \emph {et~al.}(2021)\citenamefont
  {d'Agnolo}, \citenamefont {Grosso}, \citenamefont {Pierini}, \citenamefont
  {Wulzer},\ and\ \citenamefont {Zanetti}}]{dAgnolo:2021aun}%
  \BibitemOpen
  \bibfield  {author} {\bibinfo {author} {\bibfnamefont {Raffaele~Tito}\
  \bibnamefont {d'Agnolo}}, \bibinfo {author} {\bibfnamefont {Gaia}\
  \bibnamefont {Grosso}}, \bibinfo {author} {\bibfnamefont {Maurizio}\
  \bibnamefont {Pierini}}, \bibinfo {author} {\bibfnamefont {Andrea}\
  \bibnamefont {Wulzer}}, \ and\ \bibinfo {author} {\bibfnamefont {Marco}\
  \bibnamefont {Zanetti}},\ }\bibfield  {title} {\enquote {\bibinfo {title}
  {{Learning New Physics from an Imperfect Machine}},}\ }\href@noop {} {\
  (\bibinfo {year} {2021})},\ \Eprint {http://arxiv.org/abs/2111.13633}
  {arXiv:2111.13633 [hep-ph]} \BibitemShut {NoStop}%
\bibitem [{\citenamefont {Nachman}\ and\ \citenamefont
  {Shih}(2020)}]{Nachman:2020lpy}%
  \BibitemOpen
  \bibfield  {author} {\bibinfo {author} {\bibfnamefont {Benjamin}\
  \bibnamefont {Nachman}}\ and\ \bibinfo {author} {\bibfnamefont {David}\
  \bibnamefont {Shih}},\ }\bibfield  {title} {\enquote {\bibinfo {title}
  {{Anomaly Detection with Density Estimation}},}\ }\href {\doibase
  10.1103/PhysRevD.101.075042} {\bibfield  {journal} {\bibinfo  {journal}
  {Phys. Rev. D}\ }\textbf {\bibinfo {volume} {101}},\ \bibinfo {pages}
  {075042} (\bibinfo {year} {2020})},\ \Eprint
  {http://arxiv.org/abs/2001.04990} {arXiv:2001.04990 [hep-ph]} \BibitemShut
  {NoStop}%
\bibitem [{\citenamefont {Andreassen}\ \emph {et~al.}(2020)\citenamefont
  {Andreassen}, \citenamefont {Nachman},\ and\ \citenamefont
  {Shih}}]{Andreassen:2020nkr}%
  \BibitemOpen
  \bibfield  {author} {\bibinfo {author} {\bibfnamefont {Anders}\ \bibnamefont
  {Andreassen}}, \bibinfo {author} {\bibfnamefont {Benjamin}\ \bibnamefont
  {Nachman}}, \ and\ \bibinfo {author} {\bibfnamefont {David}\ \bibnamefont
  {Shih}},\ }\bibfield  {title} {\enquote {\bibinfo {title} {{Simulation
  Assisted Likelihood-free Anomaly Detection}},}\ }\href {\doibase
  10.1103/PhysRevD.101.095004} {\bibfield  {journal} {\bibinfo  {journal}
  {Phys. Rev. D}\ }\textbf {\bibinfo {volume} {101}},\ \bibinfo {pages}
  {095004} (\bibinfo {year} {2020})},\ \Eprint
  {http://arxiv.org/abs/2001.05001} {arXiv:2001.05001 [hep-ph]} \BibitemShut
  {NoStop}%
\bibitem [{\citenamefont {Stein}\ \emph {et~al.}(2020)\citenamefont {Stein},
  \citenamefont {Seljak},\ and\ \citenamefont {Dai}}]{Stein:2020rou}%
  \BibitemOpen
  \bibfield  {author} {\bibinfo {author} {\bibfnamefont {George}\ \bibnamefont
  {Stein}}, \bibinfo {author} {\bibfnamefont {Uros}\ \bibnamefont {Seljak}}, \
  and\ \bibinfo {author} {\bibfnamefont {Biwei}\ \bibnamefont {Dai}},\
  }\bibfield  {title} {\enquote {\bibinfo {title} {{Unsupervised
  in-distribution anomaly detection of new physics through conditional density
  estimation}},}\ }in\ \href@noop {} {\emph {\bibinfo {booktitle} {{34th
  Conference on Neural Information Processing Systems}}}}\ (\bibinfo {year}
  {2020})\ \Eprint {http://arxiv.org/abs/2012.11638} {arXiv:2012.11638 [cs.LG]}
  \BibitemShut {NoStop}%
\bibitem [{\citenamefont {Benkendorfer}\ \emph {et~al.}(2020)\citenamefont
  {Benkendorfer}, \citenamefont {Pottier},\ and\ \citenamefont
  {Nachman}}]{1815227}%
  \BibitemOpen
  \bibfield  {author} {\bibinfo {author} {\bibfnamefont {Kees}\ \bibnamefont
  {Benkendorfer}}, \bibinfo {author} {\bibfnamefont {Luc~Le}\ \bibnamefont
  {Pottier}}, \ and\ \bibinfo {author} {\bibfnamefont {Benjamin}\ \bibnamefont
  {Nachman}},\ }\bibfield  {title} {\enquote {\bibinfo {title}
  {{Simulation-Assisted Decorrelation for Resonant Anomaly Detection}},}\
  }\href@noop {} {\  (\bibinfo {year} {2020})},\ \Eprint
  {http://arxiv.org/abs/2009.02205} {arXiv:2009.02205 [hep-ph]} \BibitemShut
  {NoStop}%
\bibitem [{\citenamefont {Hallin}\ \emph {et~al.}(2021)\citenamefont {Hallin},
  \citenamefont {Isaacson}, \citenamefont {Kasieczka}, \citenamefont {Krause},
  \citenamefont {Nachman}, \citenamefont {Quadfasel}, \citenamefont
  {Schlaffer}, \citenamefont {Shih},\ and\ \citenamefont
  {Sommerhalder}}]{Hallin:2021wme}%
  \BibitemOpen
  \bibfield  {author} {\bibinfo {author} {\bibfnamefont {Anna}\ \bibnamefont
  {Hallin}}, \bibinfo {author} {\bibfnamefont {Joshua}\ \bibnamefont
  {Isaacson}}, \bibinfo {author} {\bibfnamefont {Gregor}\ \bibnamefont
  {Kasieczka}}, \bibinfo {author} {\bibfnamefont {Claudius}\ \bibnamefont
  {Krause}}, \bibinfo {author} {\bibfnamefont {Benjamin}\ \bibnamefont
  {Nachman}}, \bibinfo {author} {\bibfnamefont {Tobias}\ \bibnamefont
  {Quadfasel}}, \bibinfo {author} {\bibfnamefont {Matthias}\ \bibnamefont
  {Schlaffer}}, \bibinfo {author} {\bibfnamefont {David}\ \bibnamefont {Shih}},
  \ and\ \bibinfo {author} {\bibfnamefont {Manuel}\ \bibnamefont
  {Sommerhalder}},\ }\bibfield  {title} {\enquote {\bibinfo {title}
  {{Classifying Anomalies THrough Outer Density Estimation (CATHODE)}},}\
  }\href@noop {} {\  (\bibinfo {year} {2021})},\ \Eprint
  {http://arxiv.org/abs/2109.00546} {arXiv:2109.00546 [hep-ph]} \BibitemShut
  {NoStop}%
\bibitem [{\citenamefont {Raine}\ \emph {et~al.}(2022)\citenamefont {Raine},
  \citenamefont {Klein}, \citenamefont {Sengupta},\ and\ \citenamefont
  {Golling}}]{Raine:2022hht}%
  \BibitemOpen
  \bibfield  {author} {\bibinfo {author} {\bibfnamefont {John~Andrew}\
  \bibnamefont {Raine}}, \bibinfo {author} {\bibfnamefont {Samuel}\
  \bibnamefont {Klein}}, \bibinfo {author} {\bibfnamefont {Debajyoti}\
  \bibnamefont {Sengupta}}, \ and\ \bibinfo {author} {\bibfnamefont {Tobias}\
  \bibnamefont {Golling}},\ }\bibfield  {title} {\enquote {\bibinfo {title}
  {{CURTAINs for your Sliding Window: Constructing Unobserved Regions by
  Transforming Adjacent Intervals}},}\ }\href@noop {} {\  (\bibinfo {year}
  {2022})},\ \Eprint {http://arxiv.org/abs/2203.09470} {arXiv:2203.09470
  [hep-ph]} \BibitemShut {NoStop}%
\end{thebibliography}%

\end{document}